\documentclass[twocolumn,aps]{revtex4}
\usepackage{graphicx,amsmath,epsfig} 
\usepackage{indentfirst,here}  

\begin{document} 

\title{Study of the collapse of granular columns using DEM numerical simulation}
\author{L. Staron and E. J. Hinch}%

\affiliation{Department of Applied Mathematics and Theoretical Physics, Center for Mathematical Sciences, Wilberforce Road, University of Cambridge, CB3 0WA Cambridge, UK}

\date{\today}

\begin{abstract}

Numerical simulations of the collapse and spreading of granular columns onto an  horizontal plane using the Contact Dynamics method are presented. The final shape of the deposit seems to depend only on the aspect ratio $a$ of the columns; these results are in good agreement with previous experimental work. In particular, the renormalised runout distance shows a power law dependence on the aspect ratio $a$, which is incompatible with a simple friction model. The dynamics of the collapse is shown to be mostly controlled by the free fall of the column. The energy dissipation at the base of the column can be described simply by a coefficient of restitution. Hence the energy available for the sideways flow is proportional to the initial potential energy $E_0$. The dissipation process within the flow is well approximated by basal friction, contrary to the behaviour of the runout distance. The mass ejected sideways is showned to play a determining role in the spreading process. As $a$ increases, the same fraction of initial potential energy $E_0$ drives more mass against friction. This additional dissipation give a possible explanation for power-law dependence of the runout distance on $a$. Beyond the frictional properties of the material, we show that the flow characteristics strongly depend on the early dynamics of the collapse. We propose a new scaling for the runout distance that matches the data well, is compatible with a friction model, and provide a qualitative explanation to the column collapse phenomenology.
\end{abstract}
\maketitle

\section{Introduction}
\label{intro}

Although they have been a central subject in recent research, granular flows remain intriguing in many aspects of their behaviour~\cite[see][]{rajchenbach00,goldhirsch03}. The understanding of such flows has obvious application in industrial processes, which often handle all kinds of powders and granules. But it is also crucial in the geophysical issue of catastrophic flows. Rock avalanches, landslides or pyroclastic flows all involve a granular solid phase, the influence of which can become momentous on the mean behaviour of the flow.\\
In the absence of a clear physical background for the modelling of these large scale natural granular flows, their characterisation mainly relies on the observation of the final deposits, and in particular, of the final runout distance~\cite[see][]{iverson98,dade98}. A simple effective basal friction $\mu_e$ is often advocated as the most convenient description of the dissipation process. This allows for a {\it in-situ} quantification of the mobility of the flow, knowing the runout distance and the initial height of the material. Moreover, basal friction is easily incorporated in continuous modelling, like shallow water equations for instance~\cite[see][]{savage89, mangeney04, kerswell04}. However, the dependence of this effective friction on the nature of the material, on the characteristics of the collapse or on the dynamics of the flow, has yet to be specified.\\

Systematic studies of the frictional properties and the spreading of granular flows on incline planes have been carried out~\cite[see][]{pouliquen99,pouliquen02}. However, until late, the simple case of a granular mass collapsing onto an horizontal plane had not been addressed. Recently, experimental works have tackled this problem, and studied the collapse and spreading of a suddenly released column of grains onto an horizontal plane~\cite[see][]{lube04a,lajeunesse04,balmforth04}. Essentially, the effect of the initial geometry of the column on the geometry of the final deposit has been investigated. The main result consists in scaling laws for the runout distance. In particular, when the initial aspect ratio of the collapsing column is high enough, the runout distance renormalised by the initial radius of the column shows a power law dependence on the initial aspect ratio. Moreover, this dependence varies depending on the conditions of the experiment, and in particular between the axisymmetric or quasi 2D configurations.\\
Although simple models relying on Coulomb failure analysis or on shallow water approximation for the flow have been proposed, no clear physical understanding of the granular collapse process has been achieved yet. In particular, the scaling laws obtained for the runout distance are incompatible with a simple basal friction model, and question the mode of energy dissipation occurring in the successive steps of the collapse dynamics.\\

In this context, the use of the numerical simulation to reproduce numerically the collapse of granular columns is expected to give new interesting insights in the problem. Indeed, Discrete Element Methods allow for the simulation of each grain forming the granular mass, giving access to each grain's trajectory. We can thus hope to access a picture of the collapse phenomena otherwise very difficult to obtain from experimental setup. \\
The aim of the present work is to determine which are the mechanisms controlling the spreading dynamics: fall of the column, dissipation at the base, ejection of mass sideways and dissipation within the flow. Therefore, we have applied the Contact Dynamics algorithm~\cite[see][]{moreau94,jean94} in two dimensions. The numerical procedures are explained in section~\ref{num}. After briefly recalling and commenting the experimental results obtained by other authors~\cite[see][]{lube04a,lajeunesse04,balmforth04} in section~\ref{expe}, the different regimes of spreading are qualitatively described and scaling laws are established in section~\ref{scale}. We observe a very good agreement between the experiments and the simulations. Details of the dynamics of the vertical fall and of the sideways spreading are shown in section~\ref{dynamics}. The dissipation of energy, and the transfer of energy from vertical fall to horizontal motion, are examined in section~\ref{energy}. We show that the amount of energy available for the spreading is simply proportional to the initial potential energy of the column and that basal friction is a very good approximation of the dissipation within the sideways flow. Eventually, we conclude that the ability of the column to eject the grains sideways is a major aspect of the dynamics. This leads us to question the scaling laws obtained for the runout distance as perhaps fortuitous and corresponding to a transient regime. A new empirical fit, compatible with a friction law, and qualitatively describing the collapse phenomenology, is proposed in section~\ref{new}. Summary of the results and further discussion are presented in section~\ref{conclu}. 

%-------------------------------------------------------
\section{Numerical Procedures}
\label{num}
\subsection{Simulation Method}

The numerical methods used for the simulation of granular material are known under the generic name of Discrete Element Methods. They take into account the individual existence of each discrete grain forming the media, and most usually neglect the role of the interstitial fluid filling the space between the grains. Such an approximation makes DEM methods mostly appropriate for the modelling of dry granular matter. In the absence of any dynamics induced by the surrounding fluid (the air in the present case), the behaviour of the collection of grains is entirely driven by the usual equations of motion, and the contact laws ruling the collisions between the grains.\\

The modelling of the contact phenomena consists in a set of relations between the contact force and various quantities referring to the physical and/or chemical processes taking place at the surface of the two bodies in contact. These microscopic interactions can be of different natures. They can involve for instance elastic or plastic deformation of asperities, adhesion due to Van der Waals forces, aging process... In any case, the complexity of these phenomena cannot be directly incorporated in the contact models, first due to practical reasons of numerical feasibility or efficacy, and then, because such complex models are nevertheless not assured to be realistic.  Basically, DEM methods assume small elastic deformations, with possible viscous effects, and frictional dissipation as contact phenomenology~\cite{cundall79}. \\
It is beyond the scope of this paper to give a detailed account of the numerical method used, and further informations will be found in dedicated work. Hence we will just specify the main hypothesis ruling the behaviour of the numerical grains. In the absence of a clear physical background for the modelling of contact phenomena, a possible strategy is to assume that grains are interacting through hard-core repulsion and non-smooth Coulombic friction only. This modelling choice is adopted in Contact Dynamics (CD) algorithm~\cite[see][]{moreau94, jean94} that we have applied for the present work. This implies that two grains have to touch for the contact force to be non-zero, and no distance interaction is permitted. Once a contact is formed between two grains, the latter cannot get closer, and any normal relative motion is repulsive: the grains are perfectly rigid. The microscopic coefficient of friction $\mu$ characterizes the Coulombic friction threshold $\pm \mu N$, where $N$ is the normal force at contact, and controls the friction dissipation process at grain scale. The tangential force $T$ between two grains in contact can either be below the Coulombic friction threshold, and in that case no tangential slip motion is possible, or it can be exactly equal to the Coulombic friction threshold, and in that case, slip motion will dissipate energy. These contact laws allow for the modelling of {\em multiple collisions} between the grains, namely the fact that the grains are in contact with one or more neighbours while they are undergoing collisions with other grains. However, the simple case of a binary collision is also perfectly described. A Newtonian coefficient of restitution $\rho$ is introduced, which controls the velocity of the grains after the collision knowing the velocity before the collision. This coefficient of restitution also appears in the prediction of the velocity of all the grains in the solving process.\\
These contact laws are simplistic with regards to the  microscopic reality of contact phenomena. Nevertheless, they are sufficient ingredients to reproduce the collective dynamics of a collection of grains.

\subsection{Numerical Experiments}

\begin{figure}
\begin{center}
\begin{minipage}{0.95\linewidth}
  \begin{minipage}{0.49\linewidth}
    \centerline{\includegraphics[width=0.7\linewidth]{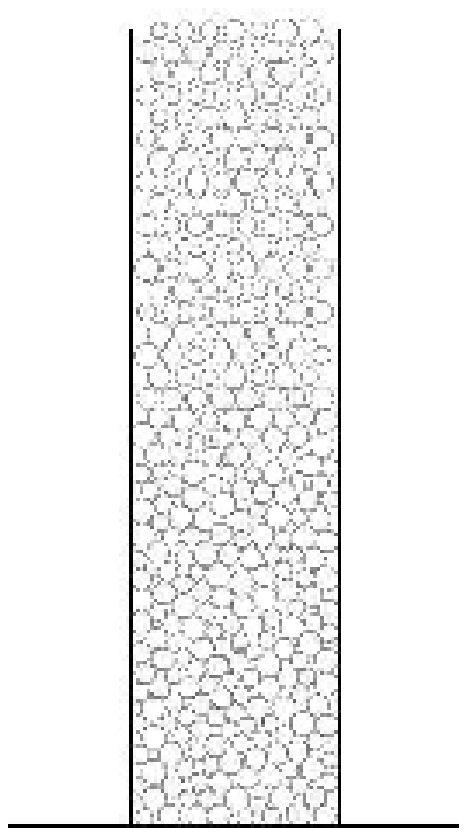}}
  \end{minipage}
  \hfill
  \begin{minipage}{0.49\linewidth}
    \centerline{\includegraphics[width=0.7\linewidth]{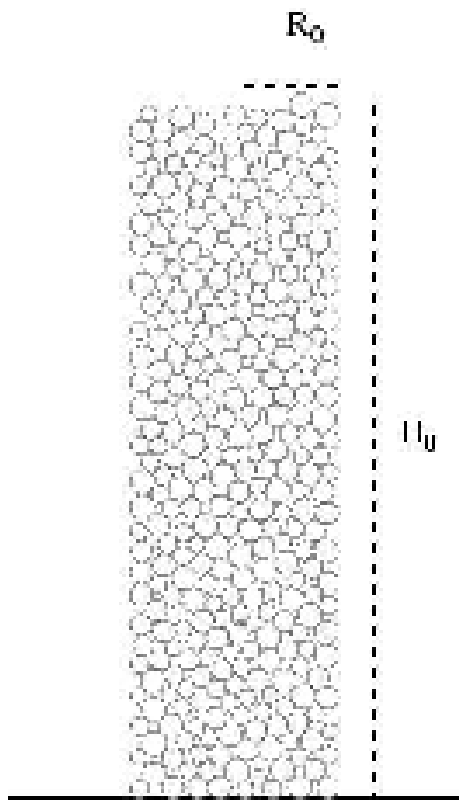}}
  \end{minipage}
  \vfill
  \begin{minipage}{0.99\linewidth}
    \centerline{\includegraphics[width=0.99\linewidth]{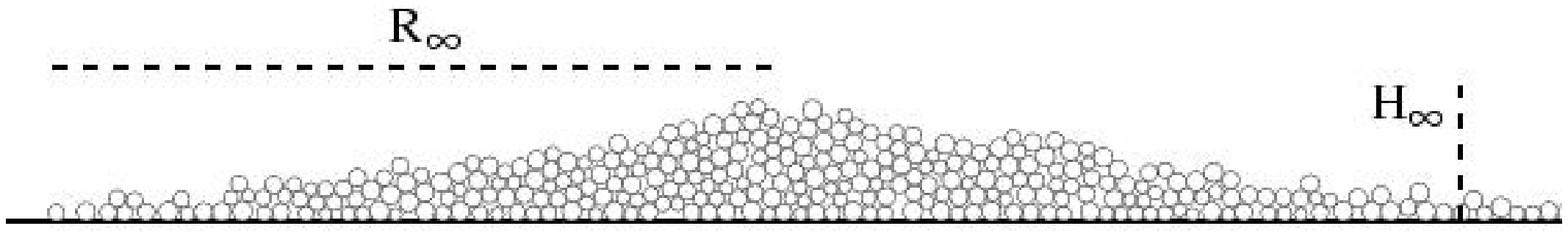}}
  \end{minipage}
\end{minipage}
\end{center}
\caption{Preparation of a column of grains by a random rain of grains in the gravity field (top left picture). The column is characterized by its initial radius $R_0$ and it initial height $H_0$. After the collapse, the final deposit is characterized by the runout distance $R_\infty$ and the height $H_\infty$ }
%\caption{times 8,23,30,80}
\label{Illus}
\end{figure}

Using the CD method, we have simulated two-dimensional collections of non-adhesive rigid circular grains. The diameter $d$ of the grains is uniformly distributed in a small interval such as $d_{min}/d_{max} = 2/3$. This slight polydispersity of grains size is chiefly introduced to break any crystal like ordering of the grains which may have a non-negligible effect in 2D simulations. In the following, $d$ denotes the mean grains diameter.\\
The coefficient of restitution $\rho$ at collision is the same between all the grains and between the grains and the bottom plane, and was chosen $\rho = 0.5$. As well, the microscopic coefficient of friction is the same at all contacts and equals $\mu = 1$. \\
%The influence of the value of these coefficients will not be addressed in the present paper.\\

The numerical experiment consists in releasing a column of grains in the gravity field onto a flat bottom plane, and study the collapse and the resulting spreading dynamics of the granular mass. The bottom plane in our experiment is perfectly smooth. Laboratory experiments carried out on rough or smooth surfaces have shown that the general behaviour of the mass of grains (in particular the final runout distance) is not affected by the roughness of the plane, save a slight transformation of the deposit final shape~\cite[see][]{lajeunesse04}.\\
The initial columns are prepared by means of a random rain of grains between two vertical walls. The compacity of the packing is $c_0\simeq 0.82$. The dimensions of the column are its radius $R_0$ and it height $H_0$, and $a = H_0/R_0$ is the initial aspect ratio. At time $t=0$, the vertical walls are instantaneously removed, and the column collapses due to gravity. The dimension of the final deposit are the final runout distance $R_\infty$, and the final height $H_\infty$. The compacity of the final deposit is $c_\infty \simeq 0.78$, namely close to the initial compacity in spite of a slight loosening of the packing of grains. This successive steps are illustrated in Figure~\ref{Illus}.

We have carried out 25 simulations, with $a$ ranging between $0.21$ and $17$, and counting from 1000 to 8000 grains.

%-----------------------------------------------

\section{Comments on the experimental results}
\label{expe}

Recently, experiments by Lube {\it et al.}~\cite[see][]{lube04a} and Lajeunesse {\it et al}~\cite[see][]{lajeunesse04} have investigated the axisymmetric collapse of a column of grains onto a flat horizontal plane. 
 The main result consists in scaling laws for the runout distance. Using the notation mentioned above (namely $H_0$ and $R_0$ are the initial height and radius of the column respectively, $a$ the initial aspect ratio, and $H_\infty$ and $R_\infty$ are respectively the height of the final deposit and the runout distance), the first authors~\cite[see][]{lube04a} find:

\[
  \frac{R_\infty-R_0}{R_0} \simeq \left\{
    \begin{array}{ll}
     1.24\:a, & a \lesssim 1.7\\[2pt]
     1.6\: a^{1/2} & a\gtrsim 1.7.
    \end{array} \right.
\]

while Lajeunesse {\it et al}~\cite[see][]{lajeunesse04} obtain:

\[
  \frac{R_\infty-R_0}{R_0} \simeq \left\{
    \begin{array}{ll}
     1.\:a, & a \lesssim 0.74\\[2pt]
     2.\: a^{1/2} & a\gtrsim 0.74.
    \end{array} \right.
\]

Moreover, quasi-2D experiments were carried out by Lube {\it et al}~\cite[see][]{lube04b} by releasing granular columns confined between two vertical walls. The following scalings were thus obtained:

\[
  \frac{R_\infty-R_0}{R_0} \simeq \left\{
    \begin{array}{ll}
     1.2\:a, & a \lesssim 2.3\\[2pt]
     1.9\: a^{2/3} & a\gtrsim 2.3.
    \end{array} \right.
\]

 Quasi-2D experiments were also carried out by Balmforth $\&$ Kerswell~\cite[see][]{balmforth04}, where the influence of the gap between the two vertical walls confining the column collapse is also addressed. The results indicates that the exponent of the power law is dependent on the size of the gap:

\[
  \frac{R_\infty-R_0}{R_0} \simeq \left\{
    \begin{array}{ll}
    \lambda a^{0.65} & $Narrow gap$ \\[2pt]
     \lambda a^{0.9} & $Large gap$
    \end{array} \right.
\]
In these experiments moreover, the prefactor $\lambda$ varies depending on the material used, while previous authors found a mono-valuated prefactor, presumably due to the narrow range of experimented material in this latter cases. \\
However, scalings found for quasi-2D experiments in a narrow gap configuration give similar results for the two authors~\cite[see][]{lube04b,balmforth04}, giving roughly ${R_\infty-R_0}{R_0} \propto a^{2/3}$. \\ 

The origin of the exponents is still open to discussion. No model yet has achieved a comprehensive explanation of the collapse dynamics. In particular, a simple friction model cannot account for them. Indeed, supposing that the initial potential energy of the column is completely dissipated by the work of the friction forces along the runout distance leads to: 
\begin{eqnarray}
\mu_e m_0 g (R_\infty-R_0) &=& m_0gH_0 ,\\
%\frac{(R_\infty-R_0)}{R_0}  &=& \frac{1}{\mu_e}\frac{H_0}{R_0},\nonumber \\
\frac{(R_\infty-R_0)}{R_0} &\propto & a^1,
\label{fric2}
\end{eqnarray}
where $m_0$ is the total mass of grains, and $\mu_e$ is the effective coefficient of friction (constant by definition). The existence of the exponents thus leads to the conclusion that the dissipation process in the collapsing columns cannot be simply interpreted as a simple basal friction, and that the overall dynamics of the spreading might be more complex than usually postulated.

%--------------------------------------------------------
%================================
\begin{figure}
%\begin{minipage}{0.98\linewidth}
  \centerline{\includegraphics[width=0.9\linewidth]{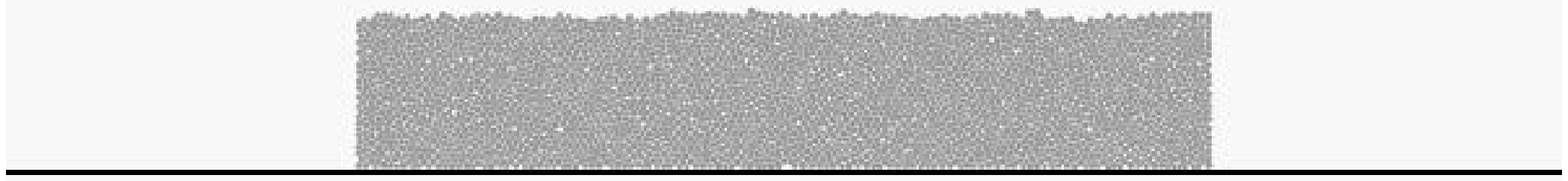}}
  \centerline{\includegraphics[width=0.9\linewidth]{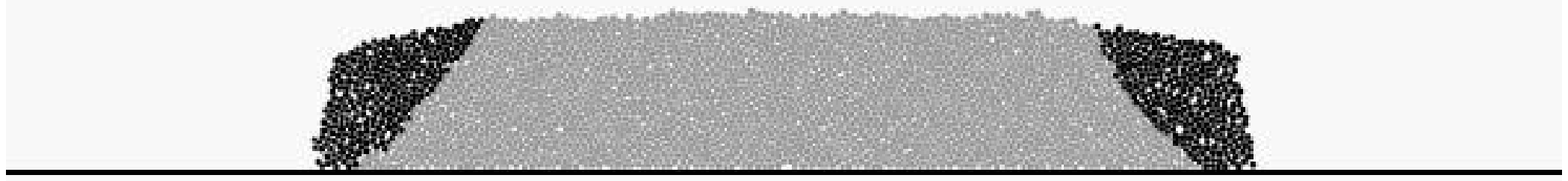}}
  \centerline{\includegraphics[width=0.9\linewidth]{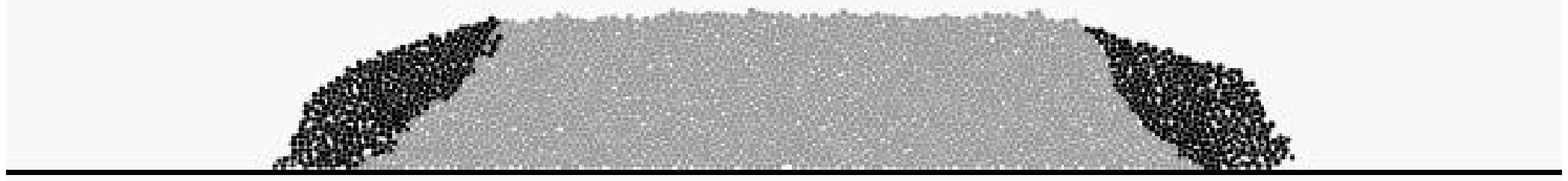}}
  \centerline{\includegraphics[width=0.9\linewidth]{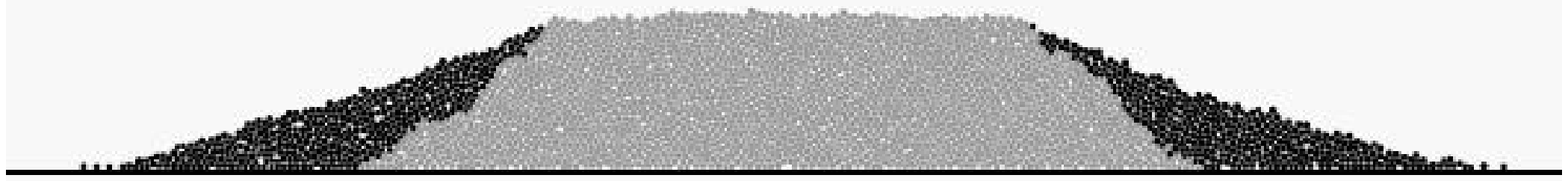}}
\caption{{\bf Grains movement in squat columns}. The snapshots show successive instants $t/T_\infty =0$, $t/T_\infty = 0.25$, $t/T_\infty = 0.37$ and  $t/T_\infty = 1$, where $T_\infty$ is the total duration of the collapse.  The column's aspect ratio is  $a=0.37$. In black are represented the grains the cumulated horizontal displacement of which exceeds the mean grains diameter $d$. The scale in the four pictures is the same.}
%{=0,8,11,infty}
\label{Fall0.37}
\end{figure}

\begin{figure}
\begin{center}
\begin{minipage}{0.9\linewidth}
  \begin{minipage}{0.49\linewidth}
    \centerline{\includegraphics[width=0.99\linewidth]{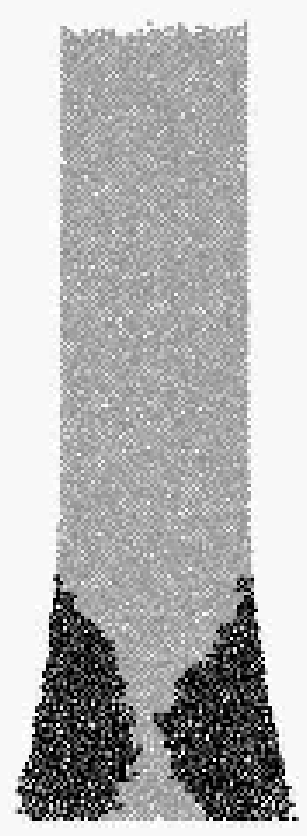}}
  \end{minipage}
  \hfill
  \begin{minipage}{0.49\linewidth}
    \centerline{\includegraphics[width=0.99\linewidth]{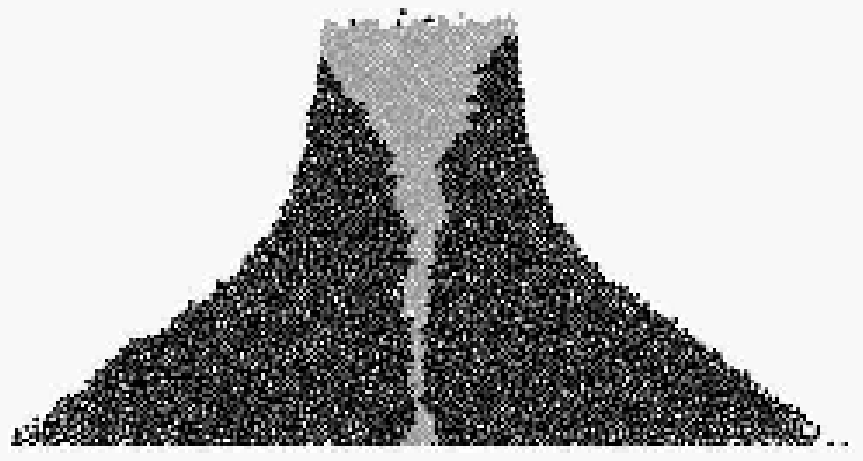}}
  \end{minipage}
  \vfill
  \begin{minipage}{0.99\linewidth}
    \centerline{\includegraphics[width=0.99\linewidth]{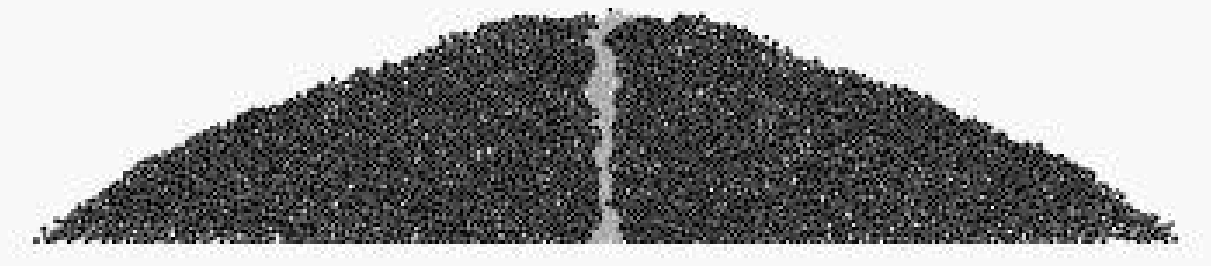}}
    \centerline{\includegraphics[width=0.99\linewidth]{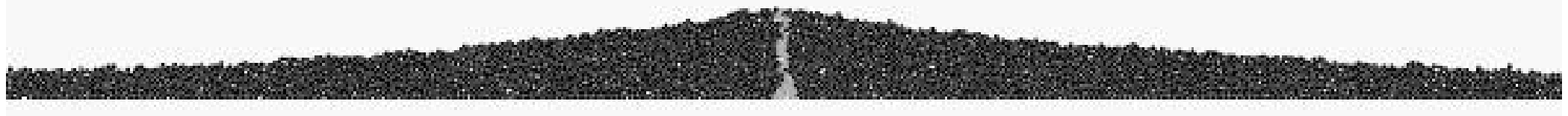}}
  \end{minipage}
\end{minipage}
\end{center}
\caption{ {\bf Grains movement in tall columns}. The snapshots show successive instants  $t/T_\infty =0.1$, $t/T_\infty = 0.25$, $t/T_\infty = 0.37$ and  $t/T_\infty = 1$, where $T_\infty$ is the total duration of the collapse.  The column's aspect ratio is  $a=9.1$. In black are represented the grains the cumulated horizontal displacement of which exceeds the mean grains diameter $d$. The scale in the four pictures is the same, and the deposit has been truncated in the last picture.}
%\caption{times 8,23,30,80}
\label{Fall9.1}
\end{figure}

\begin{figure}
    \centerline{\includegraphics[width=0.9\linewidth]{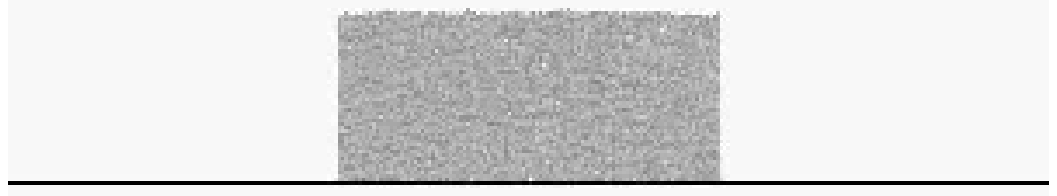}}
    \centerline{\includegraphics[width=0.9\linewidth]{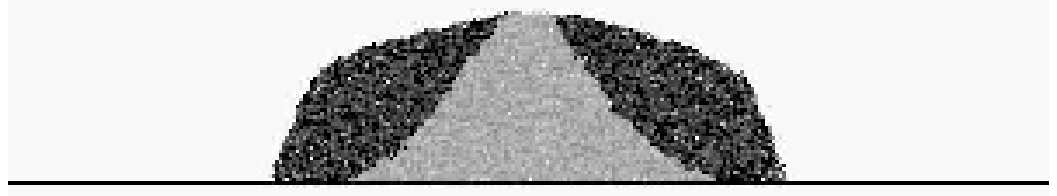}}
    \centerline{\includegraphics[width=0.9\linewidth]{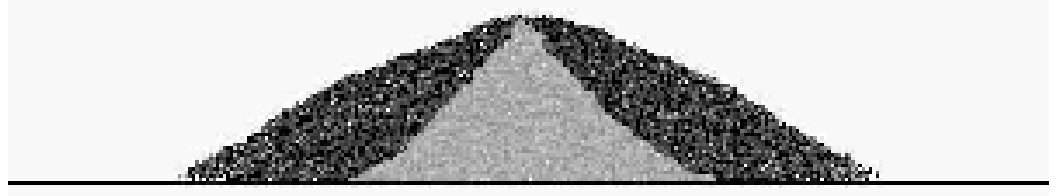}}
    \centerline{\includegraphics[width=0.9\linewidth]{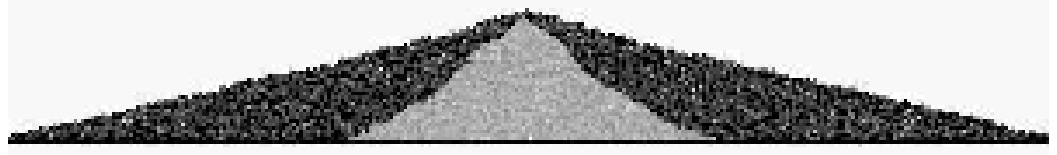}}
\caption{{\bf Grains movement in intermediate size columns}.  The snapshots show successive instants $t/T_\infty =0$, $t/T_\infty = 0.25$, $t/T_\infty = 0.37$ and  $t/T_\infty = 1$, where $T_\infty$ is the total duration of the collapse. The column's aspect ratio is  $a=0.9$. In black are represented the grains the cumulated horizontal displacement of which exceeds the mean grains diameter $d$. The scale in the four pictures is the same.}
%\caption{t=0,14,22,infty}
\label{Fall0.9}
\end{figure}

\begin{figure}
    \centerline{\includegraphics[width=0.9\linewidth]{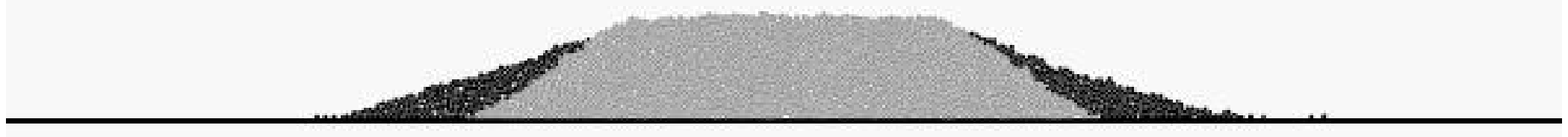}}
    \centerline{\includegraphics[width=0.9\linewidth]{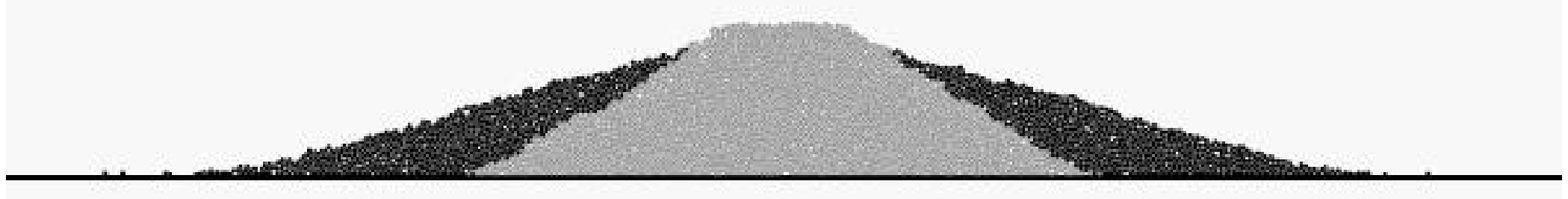}}
    \centerline{\includegraphics[width=0.9\linewidth]{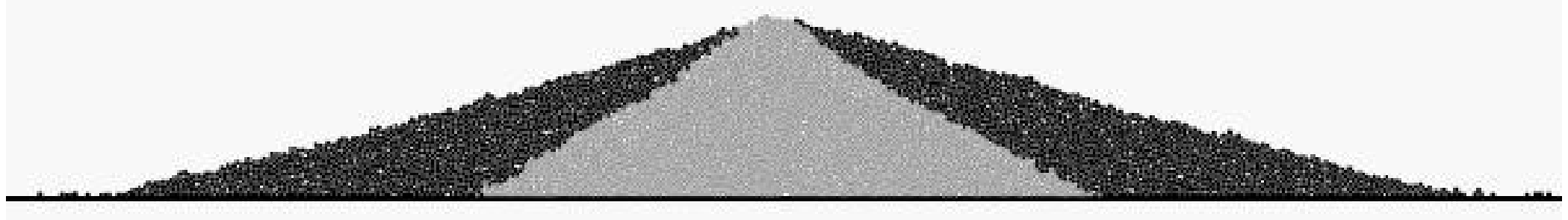}}
    \centerline{\includegraphics[width=0.9\linewidth]{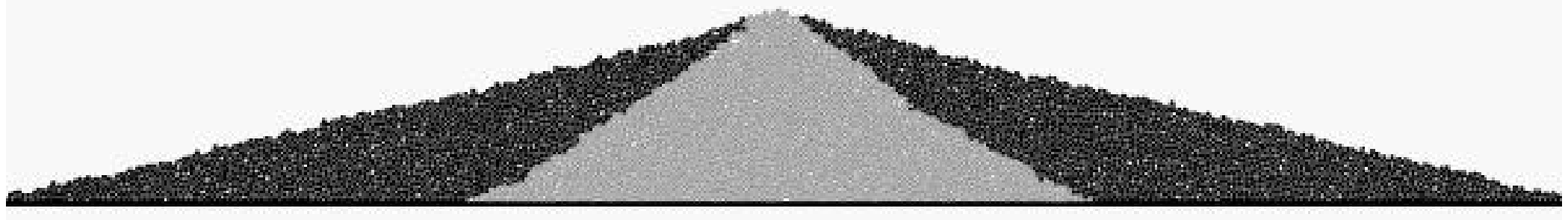}}
\caption{ {\bf The inner static cone}. Final deposits of columns with $a= 0.9$, $a= 0.73$, $a= 0.55$ and $a= 0.37$. The inner gray cone coincides with the grains whose cumulated horizontal displacement is smaller than $5d$, and shows a slope $\simeq 35^\circ$ in all cases. The scale in the four pictures is the same.}
\label{Cone}
\end{figure}
%================================================

\section{Scaling laws for the final deposit}
\label{scale}

\subsection{Qualitative Description}
\label{scaling1}

From a simple qualitative observation of the dynamics of the collapse, two different regimes can be distinguished depending on the value of the initial aspect ratio $a$. \\
In the first regime, for small $a$, the flow simply consists in the fall of the edges of the initial column. The motion is propagating from the edges inward, while a slope progressively builds up, along which the grains eventually stabilize. In this regime, only the grains situated at the sides of the columns fall and flow as a result; by contrast, the grains situated inside the column have no motion and play no role at all in the spreading. This situation is illustrated in Fig~\ref{Fall0.37}, where four snapshots of a collapsing pile with $a = 0.37$ are displayed. In the course of time, we have represented in black the grains the cumulated horizontal displacement of which exceeds the mean grains diameter $d$. We observe that a majority of grains experiences smaller or no displacement at all, and that the upper surface remains undisturbed in its larger part.\\

The second regime, namely for high $a$, is radically different. In that case, the whole column falls in a vertical motion in response to gravity, causing most of the grains to take part to the overall dynamics. Four snapshots of the collapse of a column are shown in Fig~\ref{Fall9.1} for $a = 9.1$. Again, are represented in black the grains whose cumulated horizontal displacement exceeds $d$. Only a small fraction of grains situated in the center of the column remain undisturbed. When $a$ tends to $\infty$, this fraction tends to zero.\\

An intermediate case between these two regimes is shown in Fig~\ref{Fall0.9}, for $a = 0.9$; the entire upper surface is affected by the sideways flow, but a well defined inner cone remains static. The shape of this inner cone is likely to be related to the frictional properties of the material, as suggested by~\cite{lajeunesse04}. In Fig~\ref{Cone}, final deposits are represented for columns with $a = 0.9$, $a = 0.73$, $a = 0.55$ and $a = 0.37$. To make the distinction more obvious between the flowing region and the static or creeping inner part of the column, a stronger criterion is used to distinguish the flowing grains: their cumulated horizontal displacement must exceed $5d$. By this means, the slope of the inner cone becomes very neat, and allows for the comparison of the deposits. We observe that the slope of the inner cone remains nearly constant, namely around $35^\circ$, independently of $a$. It is very tempting to assume that this slope reflects the packing mean frictional properties rather than the flow dynamics, and that the collapse, at least for small values of $a$, results from a Coulomb failure as assumed by Lajeunesse in~\cite{lajeunesse04}. In this case, assuming an hydrostatic stress gives us an internal angle of friction $\varphi \simeq 20^\circ$ for the packing of grains, equivalent to a coefficient of friction $0.36$. We note that this value does not compare with the coefficient of friction acting at contacts between grains $\mu = 1$, and neither with the effective coefficient of basal friction later evaluated in section~\ref{energy}. For greater values of $a$, the static inner cone is destroyed by the vertical dynamics of the upper grains and can no longer be observed in the final deposit.\\

The general observations on the shape of the collapsing columns just exposed are very close to previous experimental description of the collapse phenomenology.\\

\subsection{Scaling Laws for the Final Deposit}
\label{scaling2}

%=======================================================
\begin{figure}
\begin{minipage}{0.98\linewidth}
\centerline{\includegraphics[width=0.8\linewidth,angle = -90]{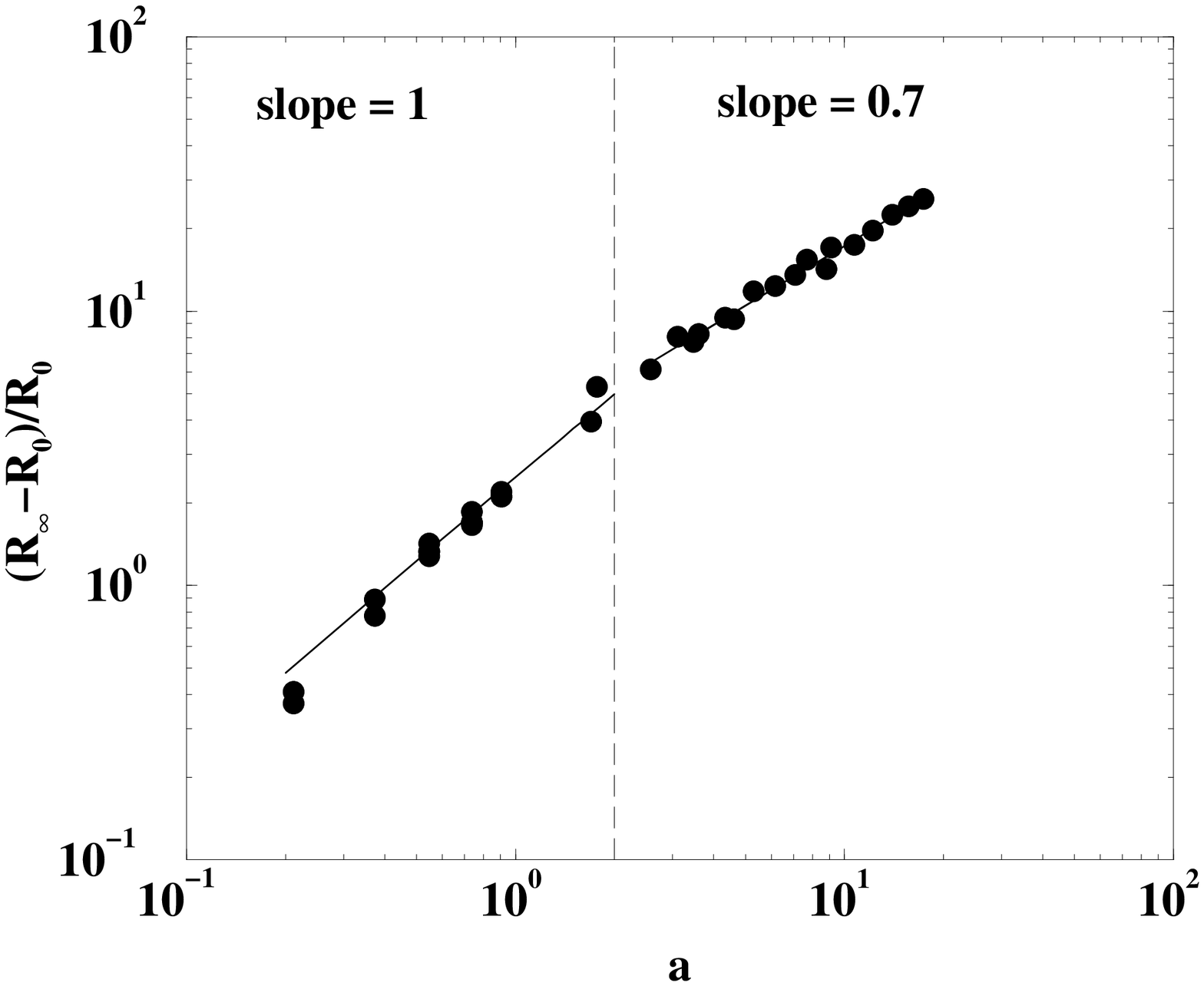}}
\caption{Final renormalised runout distance $(R_\infty-R_0)/R_0$ as a function of the columns aspect ratio $a$.\\}
\label{Rscale}
\end{minipage}
\vfill
\begin{minipage}{0.98\linewidth}
  \begin{minipage}{0.45\linewidth}
\centerline{\includegraphics[width=0.99\linewidth,angle = -90]{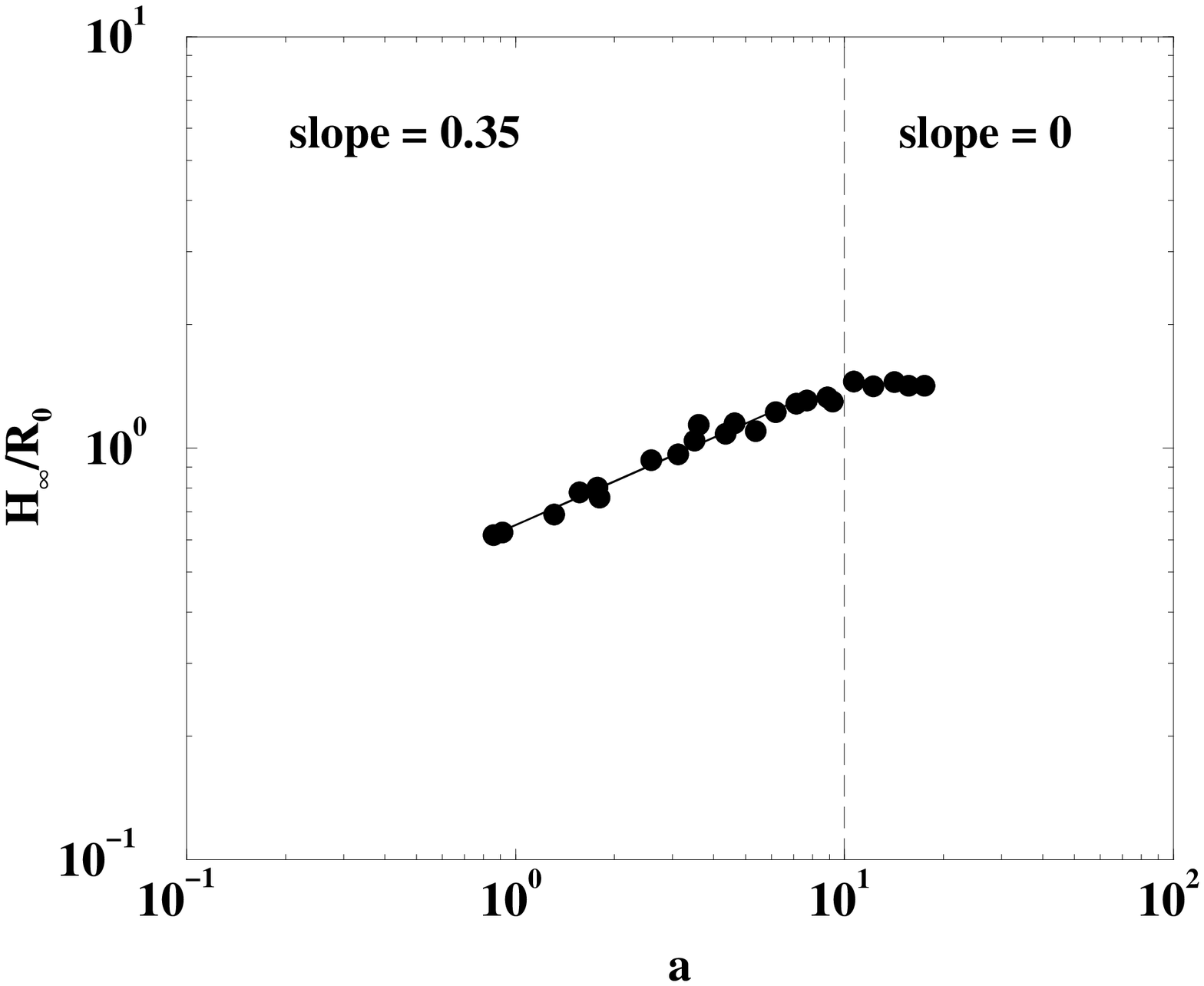}}
  \end{minipage}
  \hfill
  \begin{minipage}{0.45\linewidth}
    \centerline{\includegraphics[width=0.99\linewidth,angle = -90]{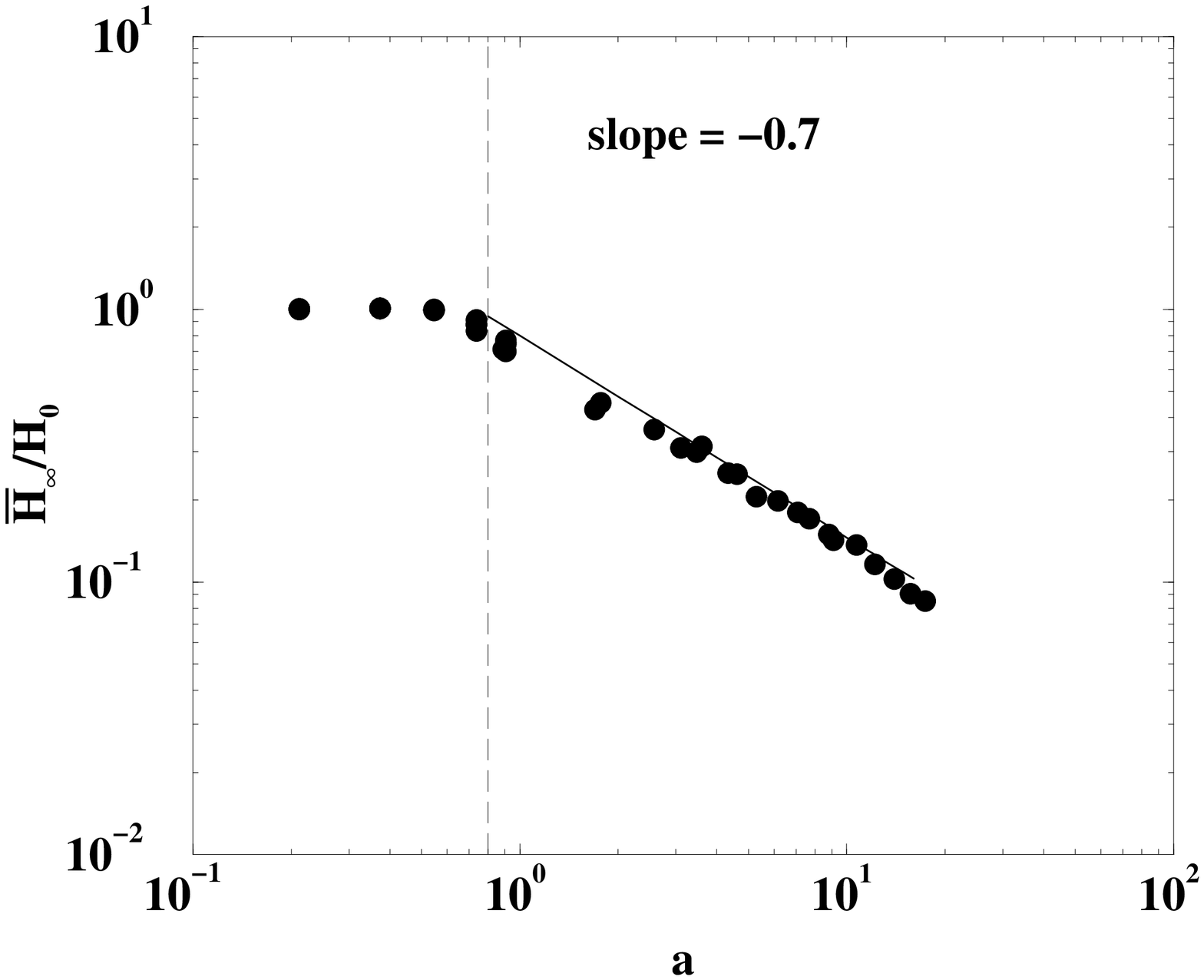}}
  \end{minipage}
\caption{Final maximum height of the deposit renormalised by the column initial  radius $H_\infty/R_0$ (left), and mean height of the final deposit renormalised by the column initial height $\overline{H}_\infty/H_0$ (right) as a function of the columns aspect ratio $a$.}
\label{Hscale}
%\end{figure}
\end{minipage}
\end{figure}

The shape of the final deposit resulting from the collapse and the spreading of the granular column is first characterized by the final runout distance $R_\infty$. For each experiment, $R_\infty$ is evaluated from the position of the grains connected to the main mass by at least one contact. In other words, grains ejected from the flow, and undergoing a solitary trajectory independently of the collective behaviour of the flow, will not be taken into account. The final height of the central conical region $H_\infty$ is also evaluated; it corresponds to the highest point of the deposit. Moreover, we evaluate the mean height of the deposit $\overline{H}_\infty$ from the total area covered by the grains. This last quantity is equivalent to a measure of the final potential energy of the deposit.\\
The evolution of the distance run by the grains normalised by the initial radius of the column $(R_\infty-R_0)/R_0$ is plotted in Fig~\ref{Rscale} as a function of $a$. We observe the following dependence:
\[
  \frac{R_\infty-R_0}{R_0} \simeq \left\{
    \begin{array}{ll}
     2.5\:a, & a \lesssim 2\\[2pt]
     3.25\:a^{0.7}, & a \gtrsim 2.
     \label{eqn1}
    \end{array} \right.
\]
As observed in laboratory experiments, the runout distance obeys two different behaviours depending on the value of $a$. For small values of $a$ a linear dependence is observed, while for larger $a$, the dependence is a power law.  The scalings we obtain are in very good agreement with the scalings observed from quasi 2D laboratory experiments, for which the exponent observed for large enough $a$ is $2/3$~\cite[see][]{lube04b,balmforth04}, and the transition between the two behaviours occurs at $a =2.3$~\cite[see][]{lube04b}.  The prefactors $2.5$ and $3.25$ obtained in the numerical experiments are higher than those observed experimentally by Lube {\it et al}, namely $1.2$ and $1.9$. This difference is very likely due to the respective frictional properties of the material; the circular shape of the numerical grains is for instance a factor of mobility. Moreover, the dissipation induced by the friction with the two vertical walls confining the grains in quasi-2D configuration is certain to affect the value of the prefactors.\\
The normalised final height $H_\infty/R_0$ is plotted against $a$ in Fig~\ref{Hscale}; we observe:

\[
  \frac{H_\infty}{R_0} \simeq \left\{
    \begin{array}{ll}
     1.6\:a^{0.35}, & a \lesssim 10\\[2pt]
     1.45, & a \gtrsim 10.
    \end{array} \right.
\]
The 2D laboratory experiments show a similar behaviour (with an exponent 0.4~\cite[see][]{lube04b} or 0.5~\cite[see][]{balmforth04}), but no transition is observed for $a\simeq 10$. \\
The increase of the quantity of grains spreading sideways appears clearly when plotting the mean height of the deposit normalized by the initial height ${\overline{H}_\infty}/{H_0}$ against the initial aspect ratio $a$. From Fig~\ref{Hscale} the dependance shows
\[
  \frac{\overline{H}_\infty}{H_0} \simeq \left\{
    \begin{array}{ll}
     1, & a \lesssim 0.8\\[2pt]
     0.4\: a^{-0.7}, & a \gtrsim 0.8,
    \end{array} \right.
\]
as can be expected from the scaling of the runout distance and the mass conservation $R_\infty \overline{H}_\infty = R_0 H_0$. This relation shows the increasing transfer of potential energy to sideways spreading motion, and the decrease with $a$ of the potential energy of the final deposit. \\
As discussed in the introduction, these scaling laws are incompatible with a simple frictional behaviour which would bring $R_\infty/R_0 \propto a$ and $\overline{H}_\infty/R_0 \propto a^{-1}$. A first hypothesis is that the dynamics of the grains at the bottom of the column is responsible for a complex dissipation process dependent on the initial aspect ratio $a$. As a consequence the energy available for the spreading would also depend on $a$, and possibly cause the dependance of $(R_\infty-R_0)/R_0$ with $a$ to be a power law. This aspect will be discussed further in section~\ref{energy}. 

%====================================================

%--------------------------------------------------------
% the dynamics

\section{Dynamics of the collapse and spreading}
\label{dynamics}

\subsection{The Vertical Fall}

%===================================
\begin{figure}
\centerline{\includegraphics[width=0.75\linewidth,angle = -90]{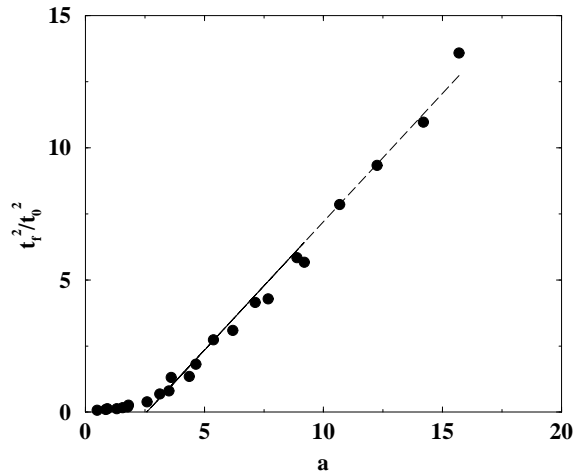}}
\caption{Dependance of $(t_f/t_0)^2$ on the column initial aspect ratio $a$, where $t_f$ is the time of free fall of the top of the column and $t_0=(2R_0/g)^{1/2}$.}
\label{FF}
\end{figure}

\begin{figure}
\begin{center}
%\begin{minipage}{0.1\linewidth}
%$t= 0.0625$\\[30pt]
%$t= 0.125$\\
%$t= 0.1875$\\
%$t= 0.250$\\
%$t= 0.3125$\\
%$t= 0.375$\\
%$t= 0.5$\\
%$t= 1.$\\
%\end{minipage}
%\hfill
\begin{minipage}{0.98\linewidth}
\begin{minipage}{0.48\linewidth}
\centerline{\includegraphics[width=0.99\linewidth]{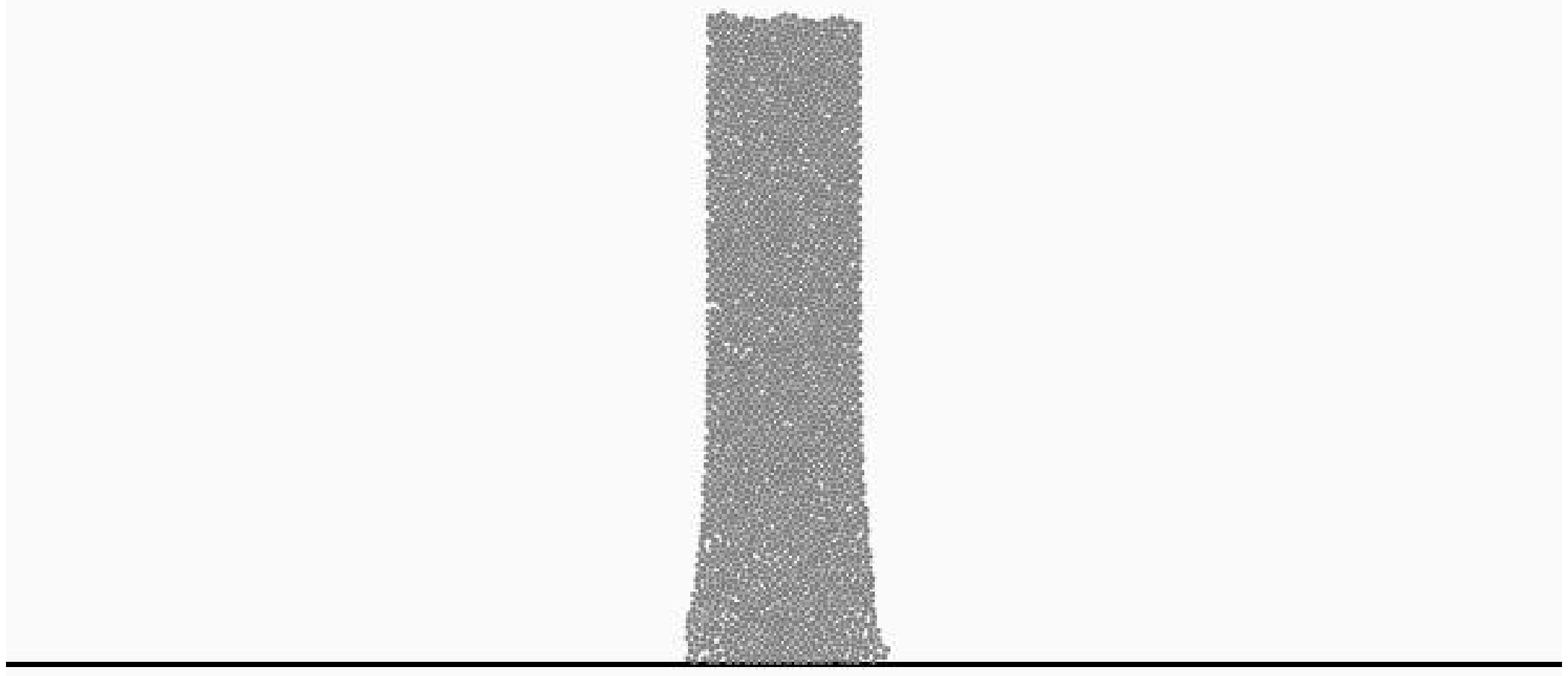}}
\centerline{\includegraphics[width=0.99\linewidth]{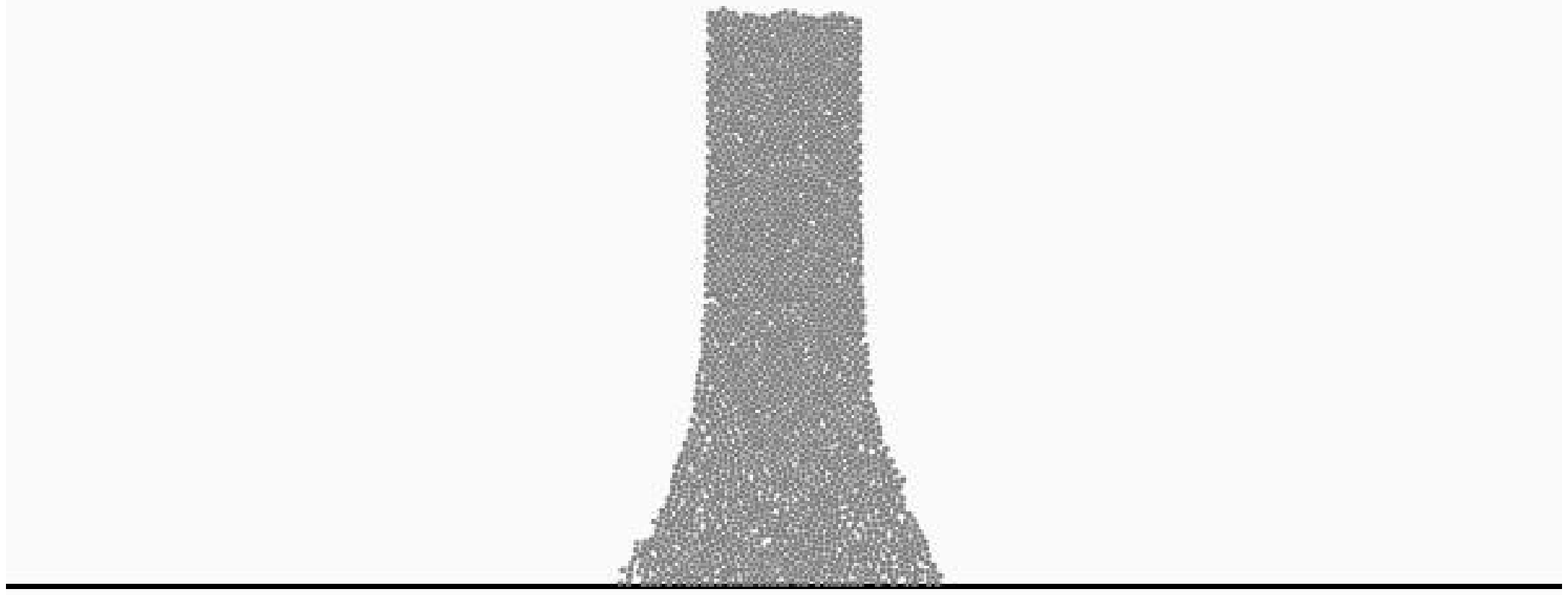}}
\centerline{\includegraphics[width=0.99\linewidth]{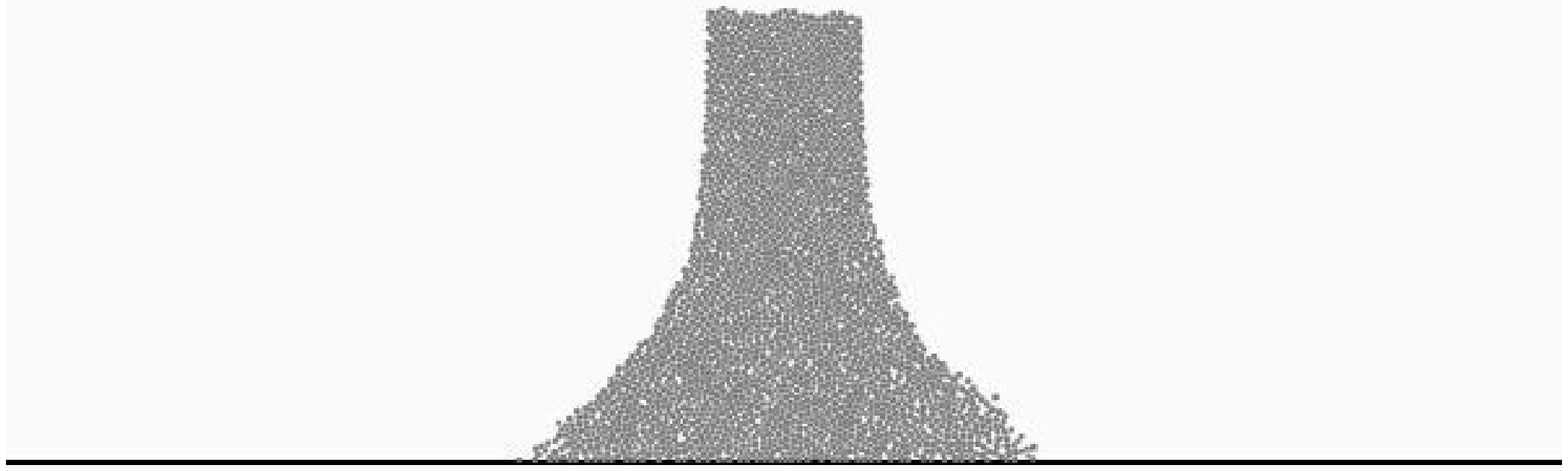}}
\centerline{\includegraphics[width=0.99\linewidth]{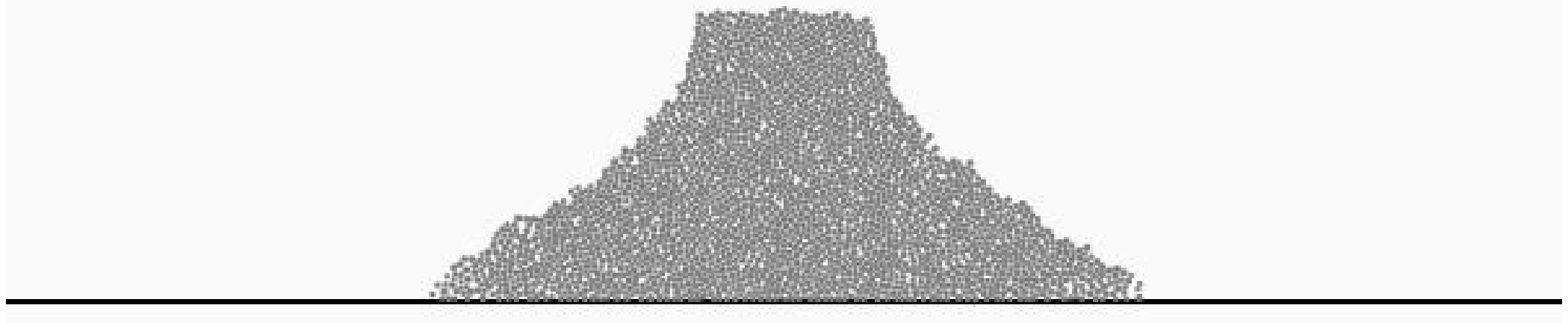}}
\centerline{\includegraphics[width=0.99\linewidth]{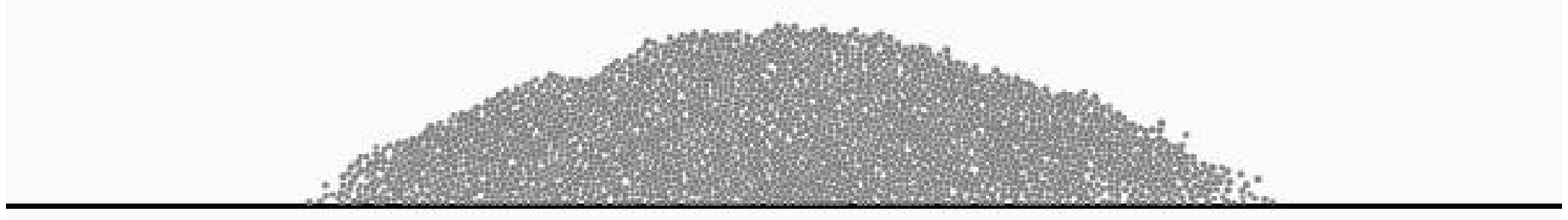}}
\centerline{\includegraphics[width=0.99\linewidth]{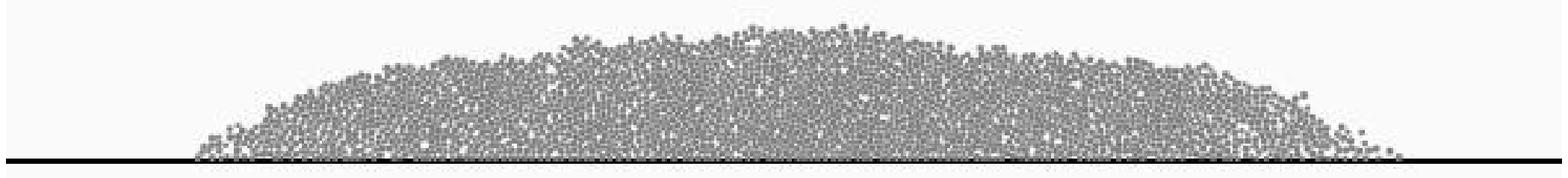}}
\centerline{\includegraphics[width=0.99\linewidth]{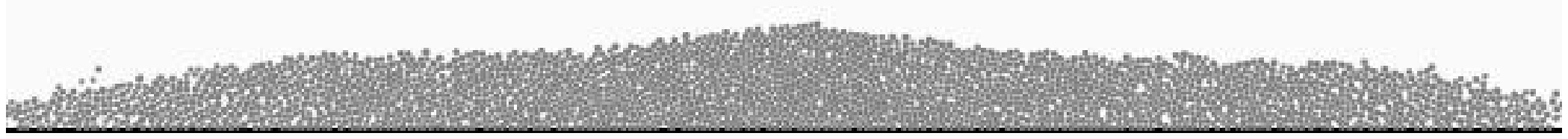}}
\centerline{\includegraphics[width=0.99\linewidth]{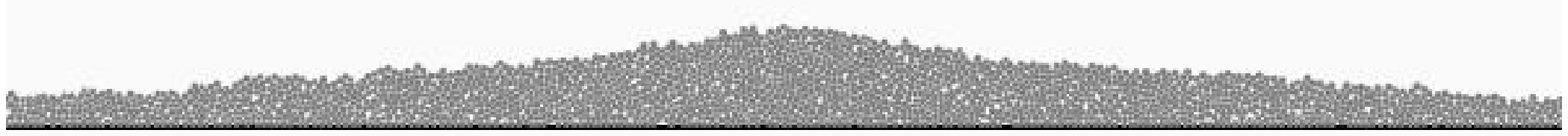}}
\end{minipage}
\hfill
\begin{minipage}{0.48\linewidth}
\centerline{\includegraphics[width=0.99\linewidth]{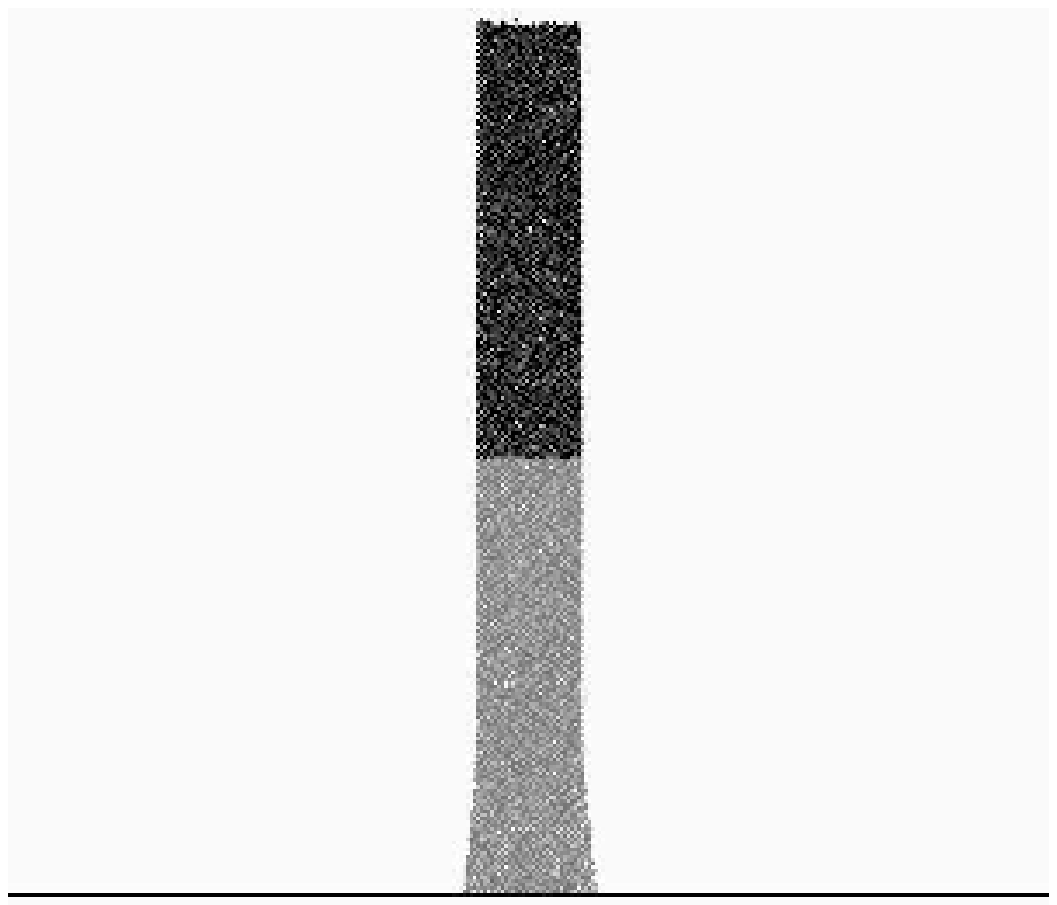}}
\centerline{\includegraphics[width=0.99\linewidth]{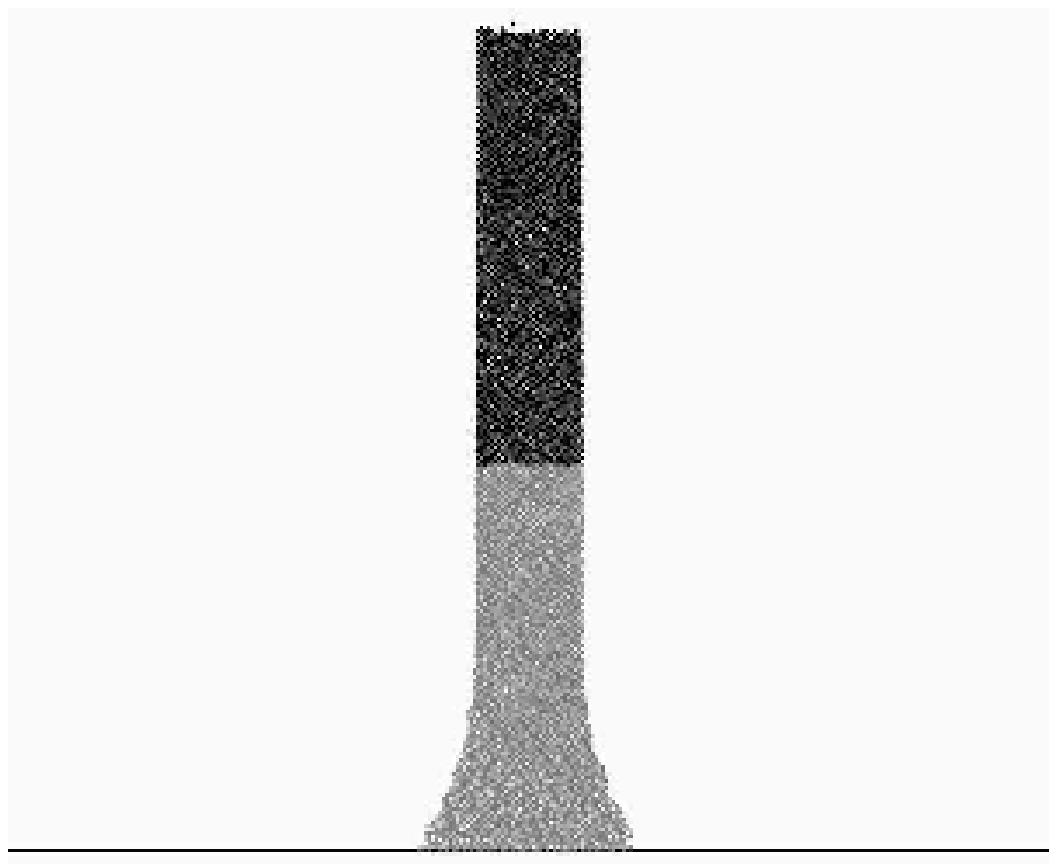}}
\centerline{\includegraphics[width=0.99\linewidth]{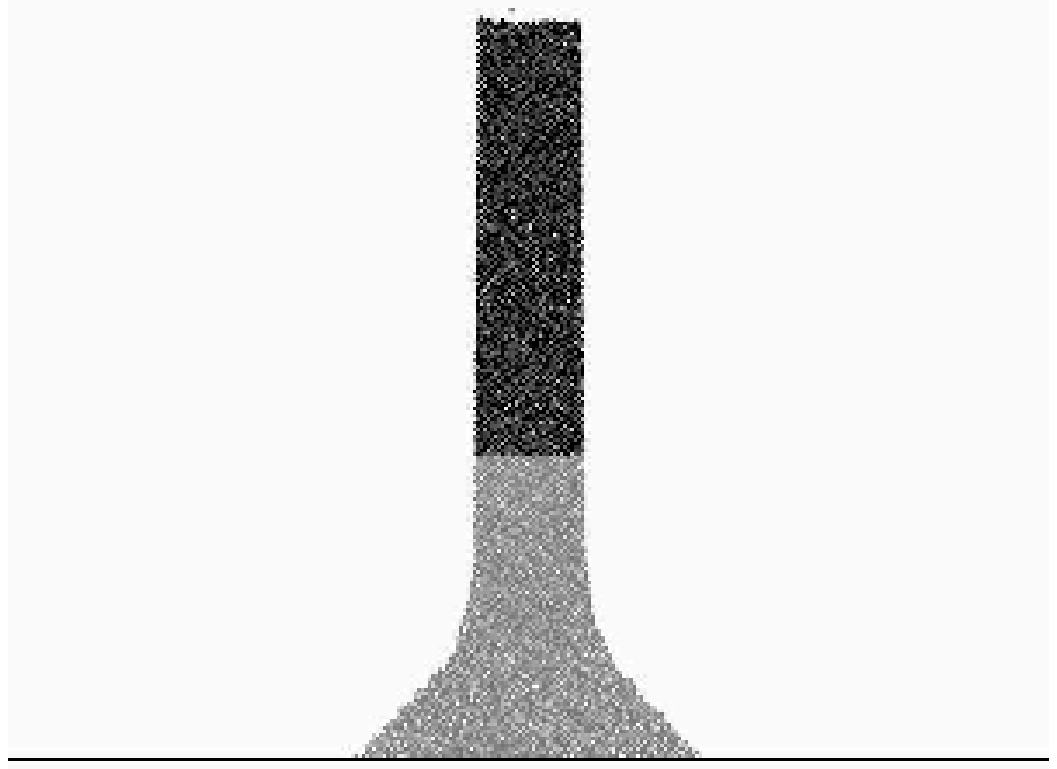}}
\centerline{\includegraphics[width=0.99\linewidth]{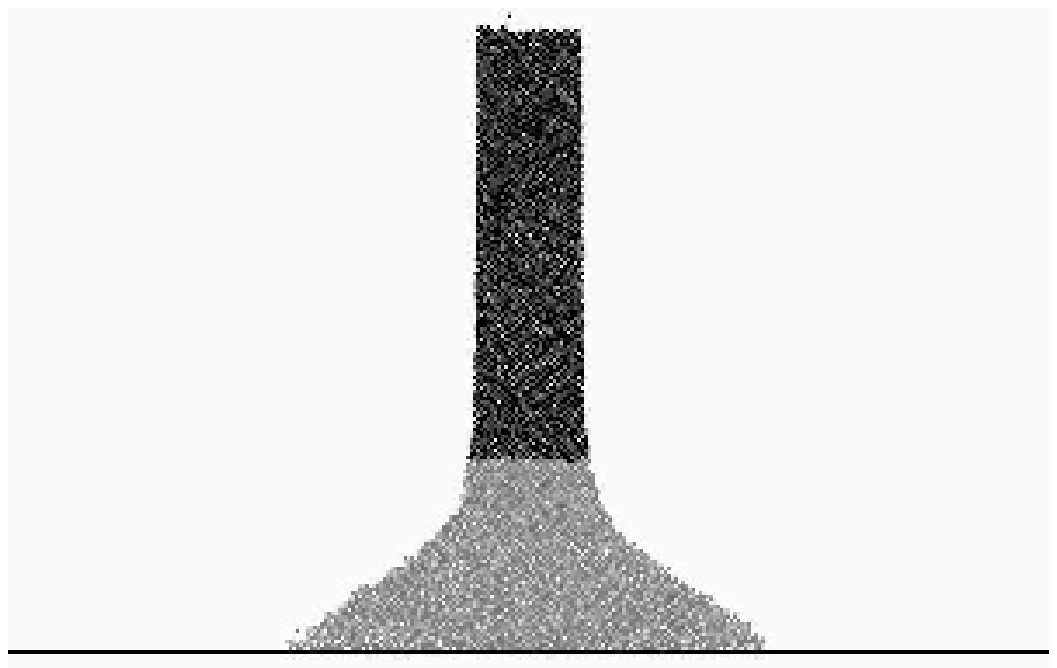}}
\centerline{\includegraphics[width=0.99\linewidth]{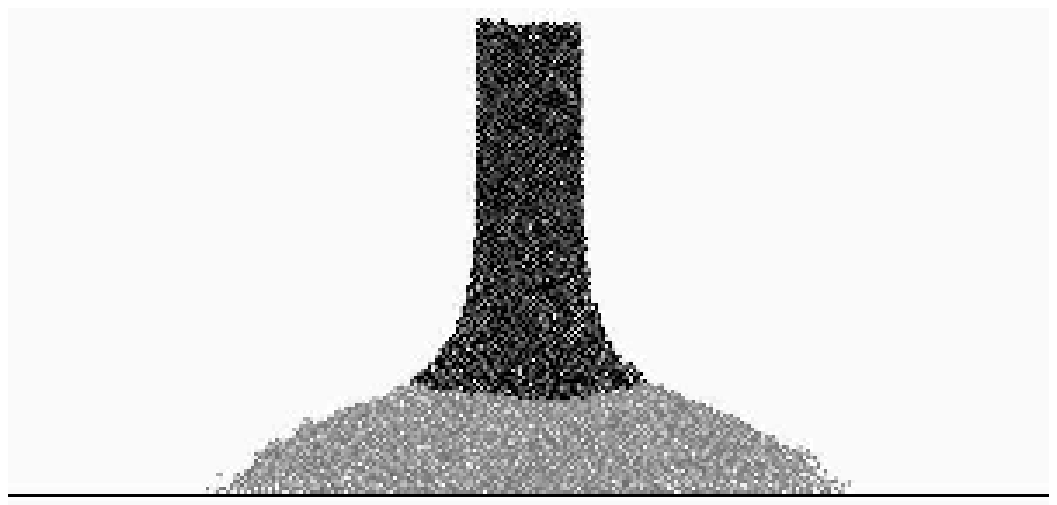}}
\centerline{\includegraphics[width=0.99\linewidth]{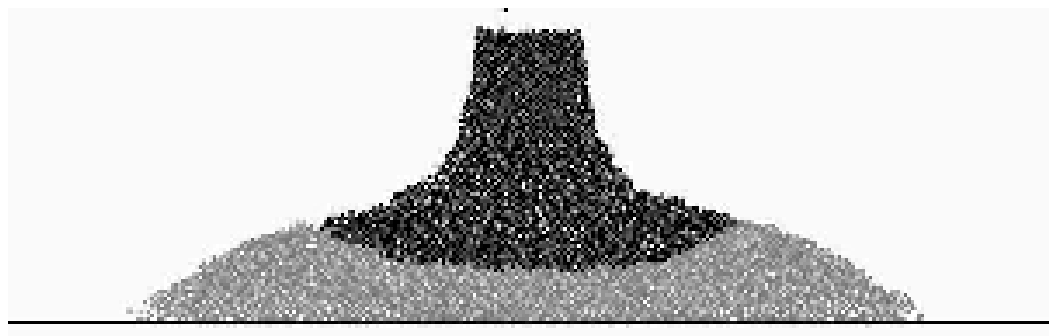}}
\centerline{\includegraphics[width=0.99\linewidth]{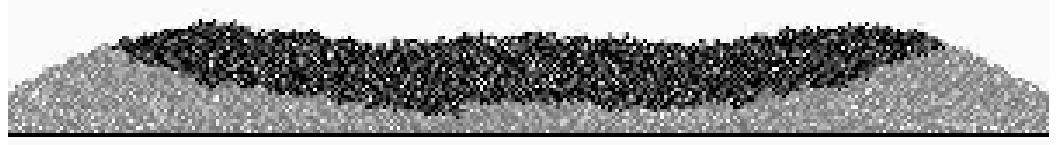}}
\centerline{\includegraphics[width=0.99\linewidth]{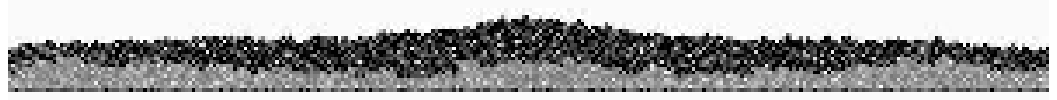}}
\end{minipage}
\end{minipage}
\end{center}
\caption{Simultaneous collapse for $a = 8.8$ and $a = 17.4$. The eight top pictures show the indifference of the spreading on the height of the columns during the free fall dynamics. The eight bottom pictures show how the additional material (in black) pushes aside and eventually covers the underlying material (in gray).}
\label{Ff}
\end{figure}

%=======================================

The dynamics of the collapse is first induced by the vertical fall of the grains. However, as stressed in section~\ref{scaling1}, different behaviour can be observed depending on the value of $a$. 
Computing the position $h$ of the top of the column in the course of time, we are able to compare $h$ with the free fall position given by $H_0 -0.5gt^2$, and to evaluate the time $t_{f}$ during which the top of the column is in free fall. The criterion used for the free fall to cease is $|h-(H_0 -0.5gt^2)|<d$ (where $d$ is the mean grains diameter).\\ 
In Fig~\ref{FF} we have plotted $(t_f/t_0)^2$ as a function of the aspect ratio $a$, where $t_0 = (2R_0/g)^{1/2}$. For small values of $a$, the time of free fall $t_f$ is nearly zero. However, when $a$ increases, the following relation is satisfied: 

\begin{eqnarray*}
\left(\frac{t_f}{t_0}\right)^2 &\simeq &0.95 a -2.5,\\
t_f &\simeq& \sqrt{\frac{2(H_0 - 2.5 R_0)}{g}}.
\end{eqnarray*}
This means that the top of the column is undergoing a free fall along a distance of nearly $H_0 - 2.5 R_0$. In other words, columns with $a \gtrsim 2.5$ are subjected to free fall, while columns with $a \lesssim 2.5$ are not. We believe this transition in the vertical dynamics of the column at $a \simeq 2.5$ to be at the origin of the transition observed in the scaling law relating $(R_\infty -R_0)/R_0$ to $a$ (see section~\ref{scaling2}, Fig~\ref{Rscale}).\\

The fact that the top of the column is in free fall entails that the upper part of the column is not affected by the complex spreading process going on at the bottom. As a consequence, two columns with the same initial radius $R_0$, but two different initial heights $H_0^1$ and $H_0^2$, should behave the same as long as the top of the smallest column remains above $\approx 2.5 R_0$. The spreading at the base should not be affected by the height of the column above, and the top of the column should no see the spreading process underneath. Once this limit height $\approx 2.5 R_0$ is reached, the smallest column will soon finish its course, while the higher column will be further accelerated.\\
This behaviour is clearly visible in the series of pictures shown is Fig~\ref{Ff}, representing the simultaneous collapse of two columns with heights $H_0^1$ and  $H_0^2 \simeq 2 H_0^1$ respectively, and same initial radius $R_0$. In the second column, the grains initially situated above the height $H_0^1$ are represented in black to allow for the comparison of the two dynamics.
From the beginning until $H_0^1 \simeq 2.5 R_0$ (first four pictures from top to bottom), the top of the two columns remain undisturbed, and the spreading process occurring at their base is exactly identical. Then, we can observe, in the respective evolution of the mass of grains initially situated under the height $H_0^1$ (represented in gray), the effect of the fall of additional grains (in black) over the underlying deposit. In particular, the dynamics of pushing aside grains that would otherwise remain in the vicinity of the bottom of the column is obvious. Eventually, the black grains cover the underlying gray ones.\\

%=======================================
\begin{figure}
\begin{minipage}[t]{0.99\linewidth}
\centerline{\includegraphics[width=0.7\linewidth,angle = -90]{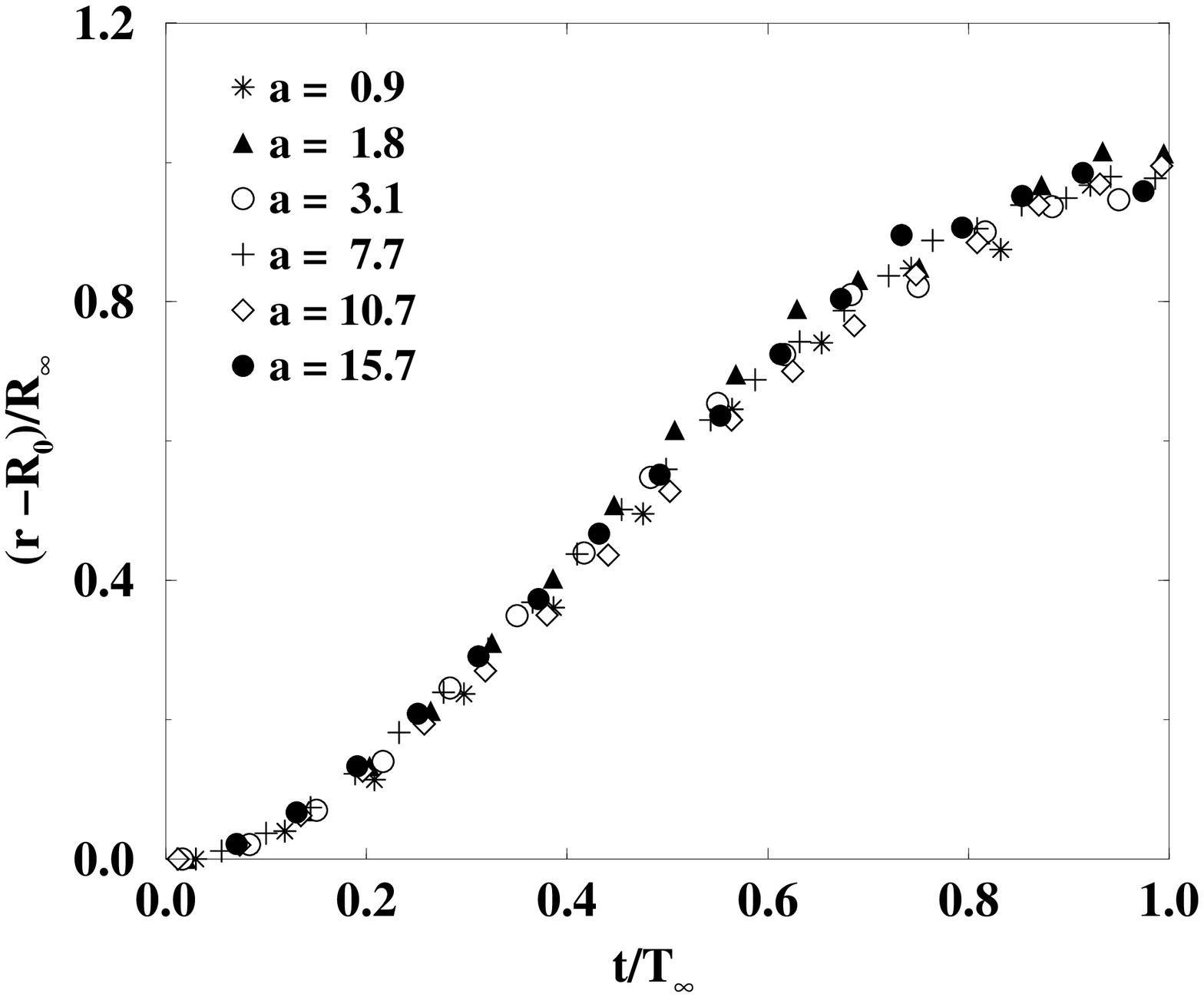}}
\caption{Front position renormalized by the final runout distance $(r-R_0)/R_\infty$ as a function of the time renormalised by the total duration of the collapse $t/T_\infty$ for different values of $a$.}
\label{Propag}
\end{minipage}
%\end{figure}
\vfill
\begin{minipage}[t]{0.99\linewidth}
%\begin{figure}
\centerline{\includegraphics[width=0.7\linewidth,angle = -90]{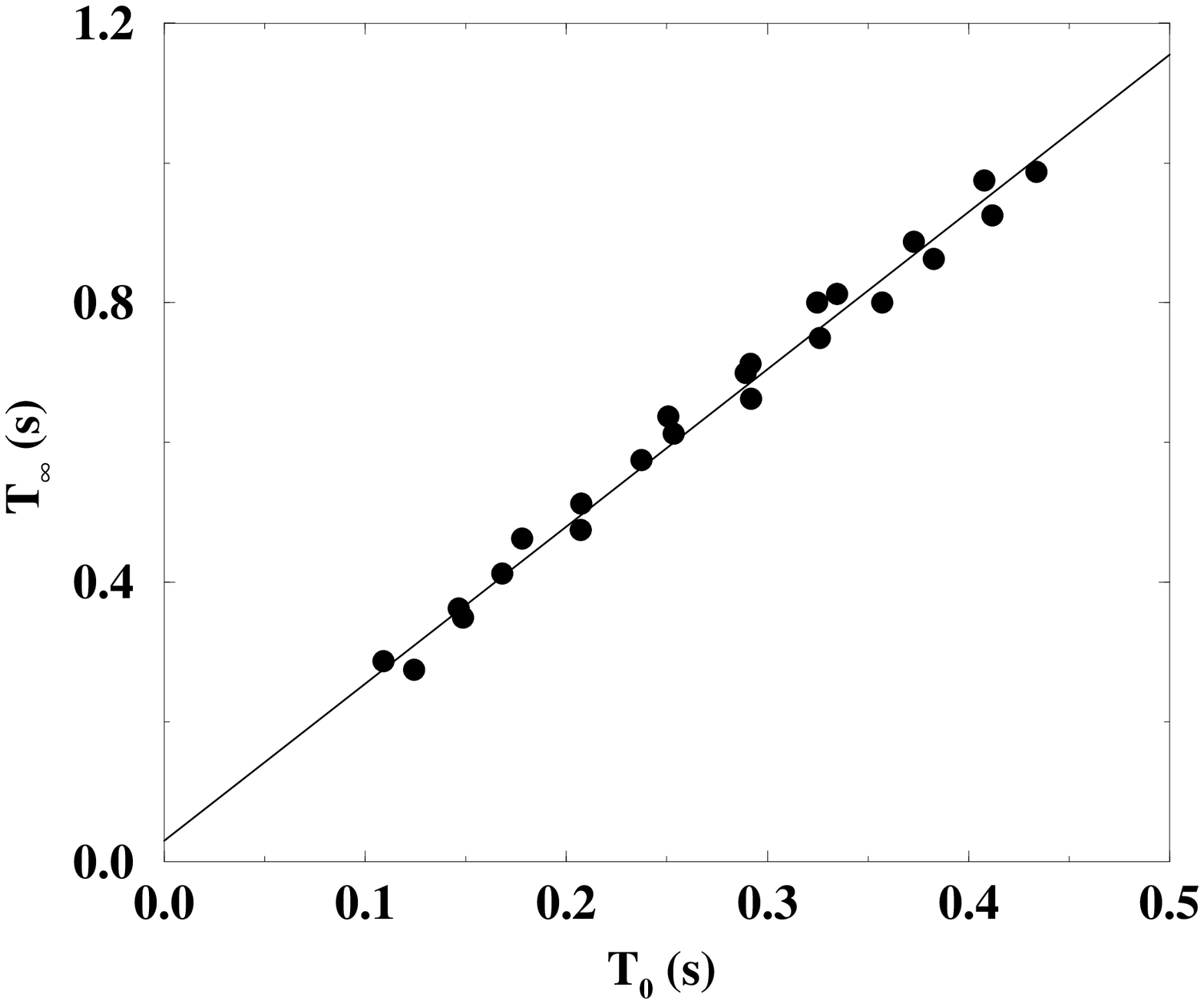}}
\caption{Total duration of the collapse $T_\infty$ as a function of $T_0 = (2gH_0)^{1/2}$. The linear relation gives $T_\infty \simeq 2.25 T_0$.}
\label{Time}
\end{minipage}
\end{figure}

\subsection{The Sideways propagation}

In the course of time, the sideways flow propagates outward; we denote $r$ the front position at any time $t$; eventually, $r$ reaches the final value $R_\infty$. The total duration of the collapse at which the sideways propagation stops and the whole deposit comes to repose is denoted $T_\infty$. \\
In order to compare the dynamics of the spreading for different values of $a$, we plot in the same graph in Figure~\ref{Propag} the evolution of the front position normalised by the final runout distance $(r-R_0)/R_\infty$ as a function of the time normalised by the total duration of the propagation $t/T_\infty$, for $a =0.9$, $a =1.8$, $a =3.1$, $a =7.7$, $a =10.7$ and $a =15.7$. The plots nicely collapse into a master curve, showing first a period of acceleration of the front, followed by a constant propagation regime, and then a deceleration period.\\ 

From the initial geometry of the column two characteristic times $t_0 =(2R_0/g)^{1/2}$ and $T_0 =(2H_0/g)^{1/2}$ can be formed, corresponding to the time of free fall along the distance $R_0$ and $H_0$ respectively. The collapse duration $T_\infty$ is plotted against $T_0$ in graph~\ref{Time}. We observe a linear relation
$$T_\infty \simeq 2.25 T_0 = 2.25 \left(\frac{2 H_0}{g}\right)^{1/2},$$
implying that the duration of the experiment is controlled by the free fall of the column of initial height $H_0$. This is in agreement with experimental results, for which the duration of the experiments is found to be nearly $3 T_0$~\cite[see][]{lube04b}.\\  

\begin{figure}
\centerline{\includegraphics[width=0.9\linewidth,angle = -90]{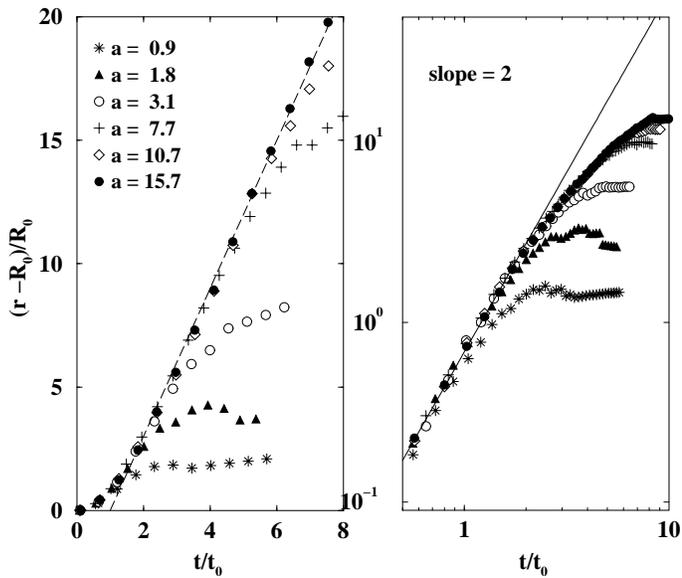}}
\caption{Front position renormalised by the column initial radius $(r-R_0)/R_0$ as a function of the renormalised time $t/t_0$ for different values of $a$ in linear (left) and logarithmic (right) representation.}
\label{Acceleration}
\end{figure}

The period of acceleration of the propagation front is well characterized when plotting for different values of $a$ the normalised front position $(r-R_0)/R_0$ as a function of the renormalised time $t/t_0$ (Fig~\ref{Acceleration}). For high values of $a$, we clearly distinguish the acceleration phase followed by a constant velocity propagation phase. This evolution is not so obvious for small values of $a$, for which the deceleration phase occurs early and leaves less time for a constant velocity regime to settle.\\
 When plotting $(r-R_0)/R_0$ against $a$ in a log-log representation, we see, up to $t/t_0 \simeq 1.5$, and for $a \gtrsim 1.8$, that the following relation is acceptable:
$$\frac{r-R_0}{R_0}\simeq 0.68\left(\frac{t}{t_0}\right)^{2}.$$
Although this approximation is made over a very short time interval, it suggests that the onset of the spreading is driven by the free fall dynamics. \\

Where a constant velocity regime can be observed, the following relation is satisfied:
$$\frac{r-R_0}{R_0} \simeq 3 \frac{t}{t_0} -3,$$ 
which is equivalent to
$${r} \simeq v_0 t,\:\:\:\: r > 2R_0.$$
where $v_0 = (2gR_0)^{1/2}$ is the front propagation velocity, once a constant velocity regime is settled, after the front position has already run a distance $2R_0$.\\
 % duscussion: parallel evolution of the front and the column top for a high a0

\begin{figure}
\centerline{\includegraphics[width=0.9\linewidth]{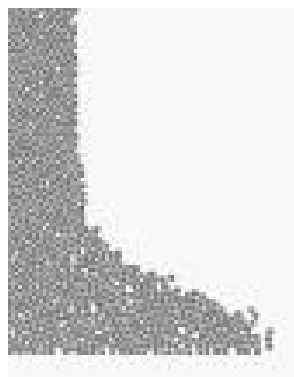}}
\centerline{\includegraphics[width=0.9\linewidth]{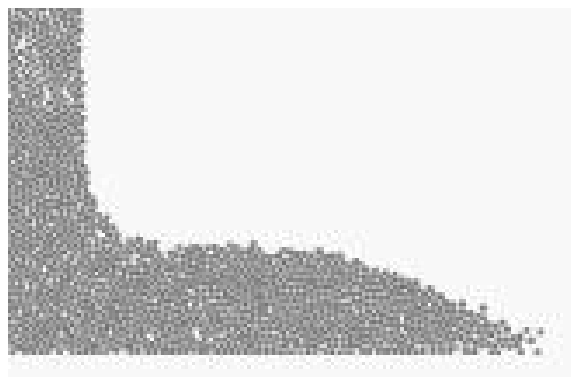}}
\centerline{\includegraphics[width=0.9\linewidth]{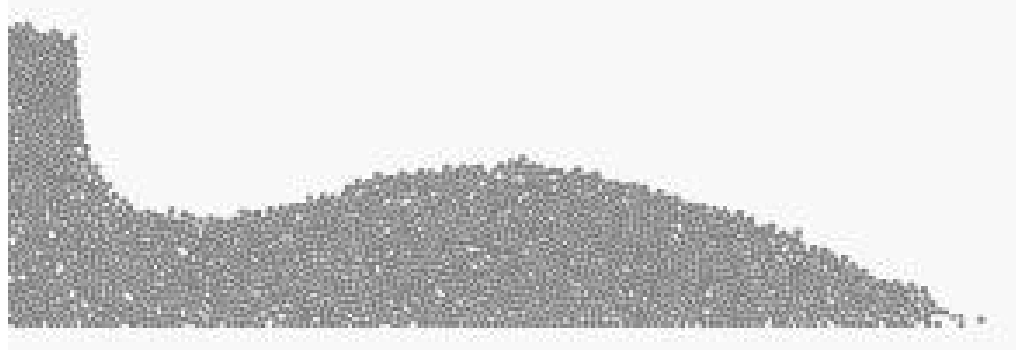}}
\centerline{\includegraphics[width=0.9\linewidth]{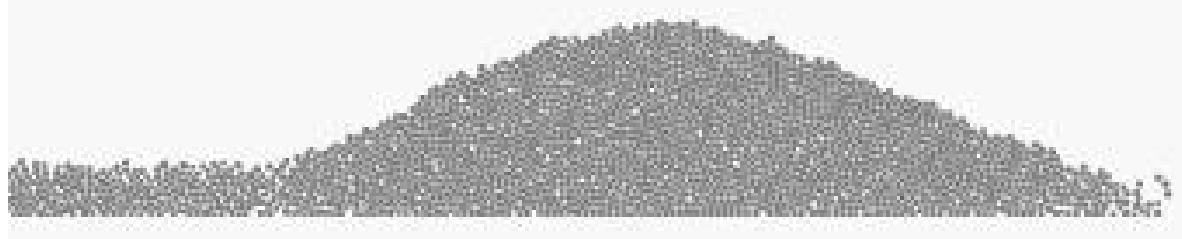}}
\centerline{\includegraphics[width=0.9\linewidth]{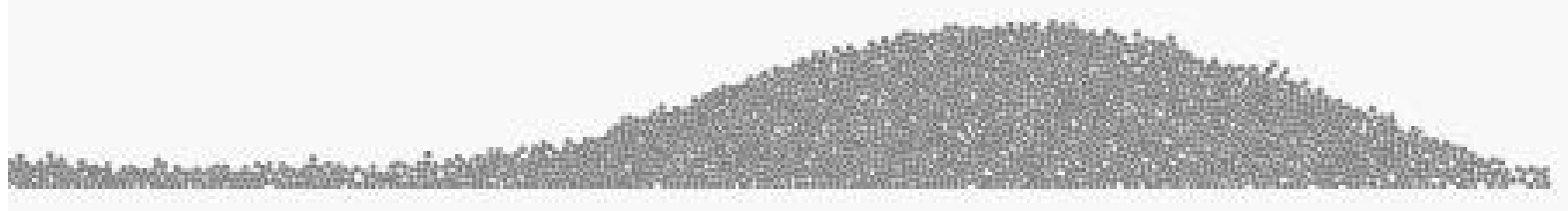}}
\centerline{\includegraphics[width=0.9\linewidth]{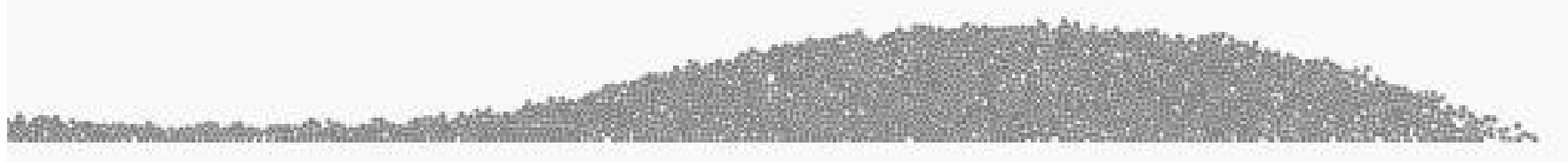}}
\centerline{\includegraphics[width=0.9\linewidth]{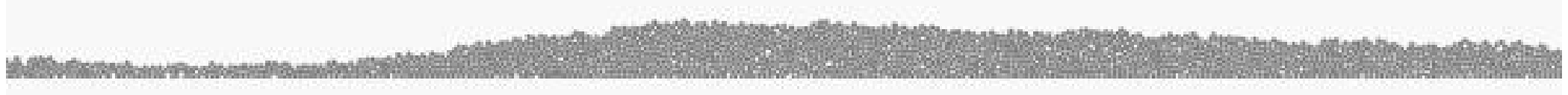}}
\caption{Successive snapshots of the spreading of a column with a very high aspect ratio $a = 70$. We observe the transfer of mass from the center of the collapse towards the front of the flow. The scale is the same on all pictures.}
%\caption{time = 14,24,38, 44,55,60, 100 (5*0.001)}
\label{Wave}
\end{figure}

Since the column is undergoing a free fall for $a \gtrsim 2$, the cumulated mass of grains $m(t)$ expelled as a result of the collapse is given by $m(t) \propto \rho_s R_0gt^2$, where $\rho_s$ is the surfacic mass of the grains packing. As the greater part of the front propagation involves a constant velocity, $r \simeq v_0 t$, the increasing mass debit at the bottom of the column can only be accommodated by an increase of the height of the sideways flow. Moreover, the grains reaching the bottom, after they have been accelerated in the gravity field, have a greater momentum than the grains preceeding them. This effect is responsible for the existence of a wave propagating outwards and transferring the mass from the center towards the margin of the spreading for high values of $a$. An extreme case of this ``mass propagation'' phenomena is illustrated in Fig~\ref{Wave} where the sideways flow is represented for $a = 70$. 

\begin{figure}
\centerline{\includegraphics[width=0.75\linewidth,angle = -90]{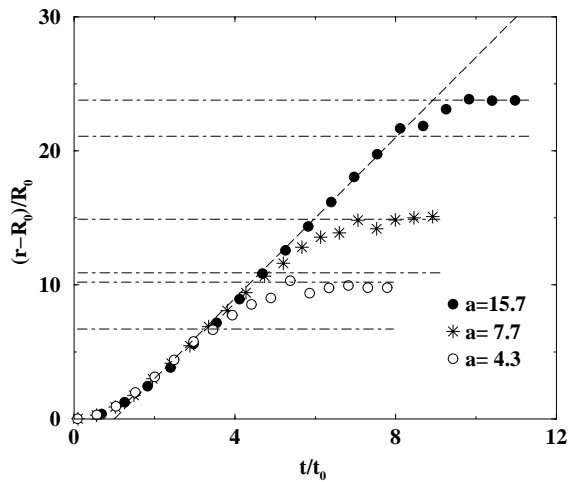}}
\caption{Front position renormalised by the column initial radius $(r-R_0)/R_0$ as a function of the renormalised time $t/t_0$ for different values of $a$. The linear evolution is stressed by the dashed line. Dashed-dotted lines indicate roughly the period of deceleration for each curve.}
\label{Decelere}
\end{figure}

This effect is more important in 2D configurations than in axisymmetric ones, for which the increase of the surface available for the spreading is quadratic with the front propagation, while the front propagation has been shown to obey the same behaviour $r = v_0 t$~\cite[see][]{lajeunesse04}. This suggests a purely geometrical explanation of the difference observed in the scaling laws in axisymmetric and 2D experiments.
Indeed, the front propagation velocity $v_0 = \sqrt(2gR_0)$, and the typical time of the experiment $T_0 = \sqrt(2H_0/g)$, give for the runout distance the straightforward scaling law:
$$(R_\infty -R_0) = v_0 T_0 = (H_0 R_0)^{1/2},$$
corresponding to the observation $(R_\infty -R_0)/R_0 \simeq a^{1/2}$ in axisymmetric collapses. The difference with the 2D case (namely $(R_\infty -R_0)/R_0 \simeq a^{2/3}$) could be due to the larger increase of mass with important momentum in the sideways flow, whose effect would be to lengthen the deceleration phase. The contribution of the deceleration phase to the runout distance is non-negligible in 2D. Three examples of the contribution of the deceleration phase to the final runout distance are displayed in Figure~\ref{Decelere}; up to one third of the total runout distance is gained during the deceleration. This effect is presumably more important than in axisymmetric configuration, for which the mass debit is more easily accommodated by the expanding surface of the flow. This could explain the difference of the exponents ($1/2$ in axisymmetric and $2/3$ in 2D) in the scaling laws. A systematic analysis of the deceleration phase should help to solve this issue; it is however not undertaken in the present paper.

%=============================================================

%==========================================================

%----------------------------------------------------
% Energy

\section{Energy transfer and dissipation}
\label{energy}

\subsection{Time evolution}

\begin{figure}
\begin{center}
\begin{minipage}{0.99\linewidth}
  \begin{minipage}{0.49\linewidth}
    \centerline{\includegraphics[width=0.99\linewidth]{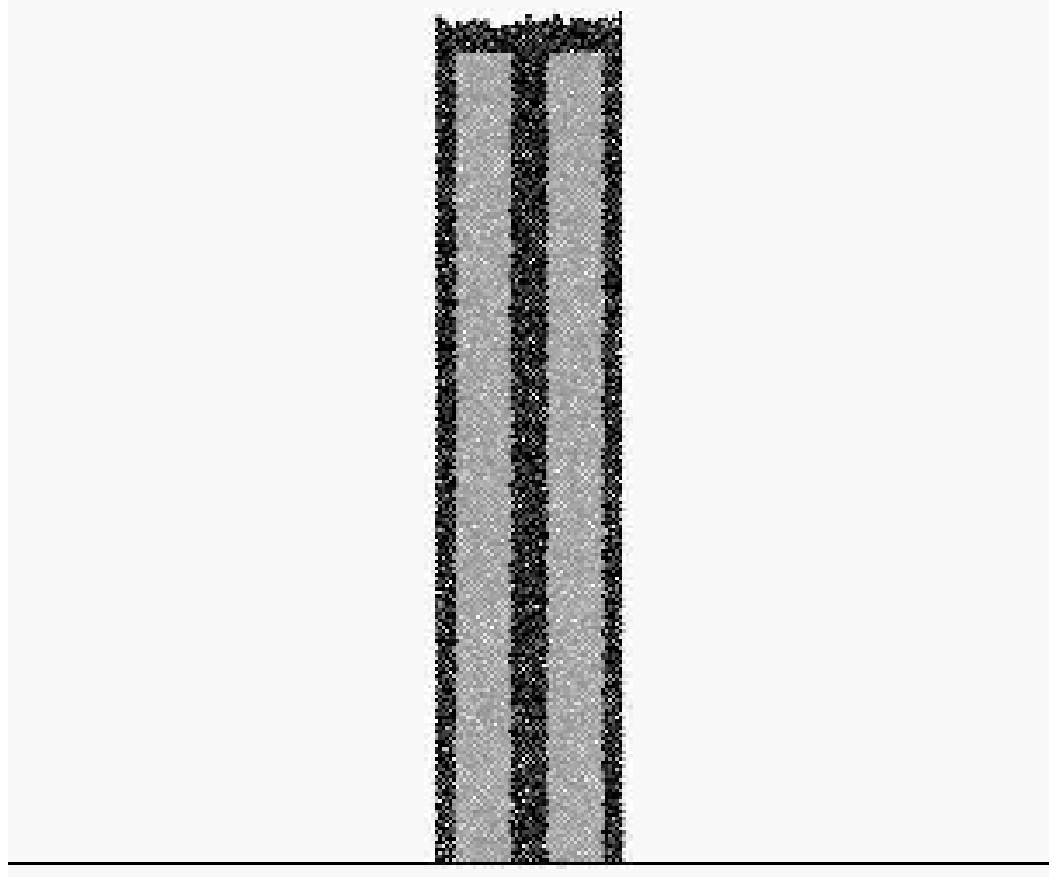}}
    \centerline{\includegraphics[width=0.99\linewidth]{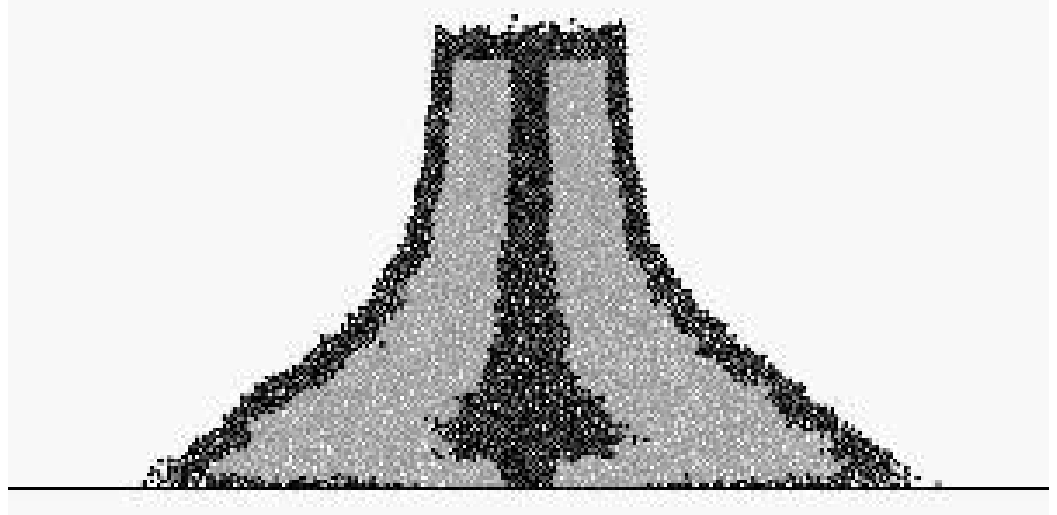}}
  \end{minipage}
  \hfill
  \begin{minipage}{0.49\linewidth}
    \centerline{\includegraphics[width=0.99\linewidth]{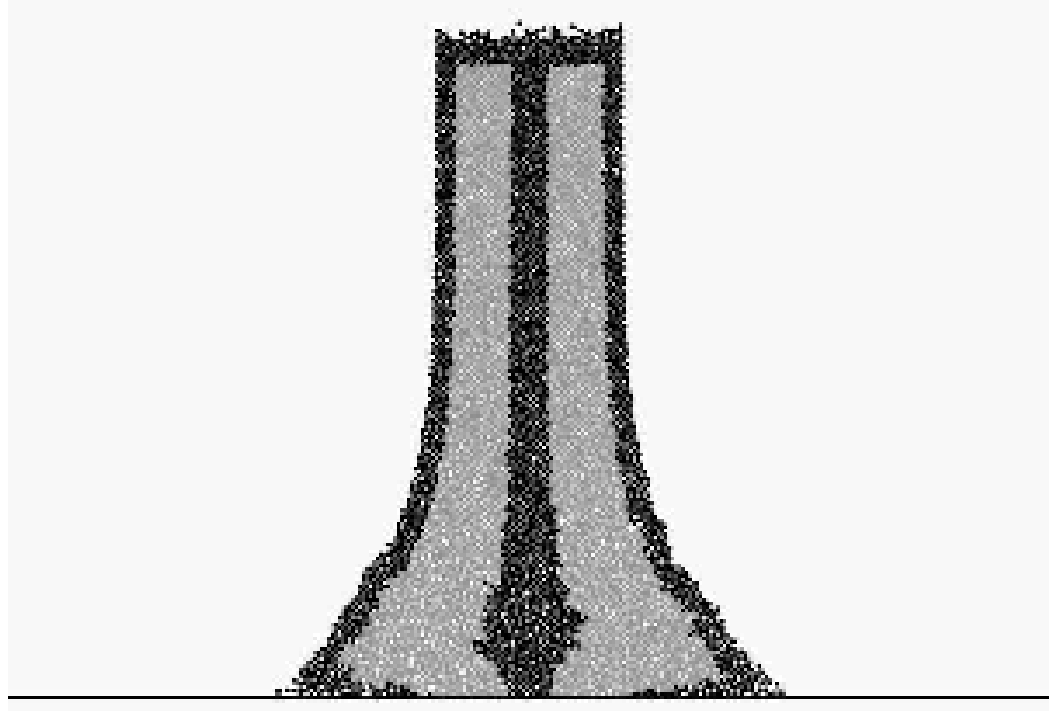}}
    \centerline{\includegraphics[width=0.99\linewidth]{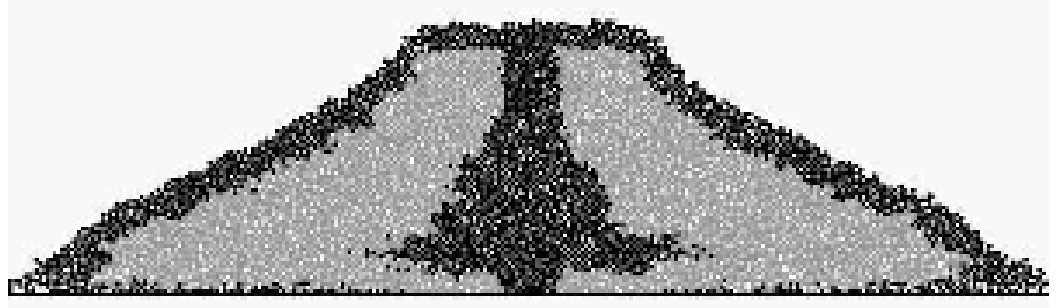}}
  \end{minipage}
  \vfill
  \begin{minipage}{0.99\linewidth}
    \centerline{\includegraphics[width=0.99\linewidth]{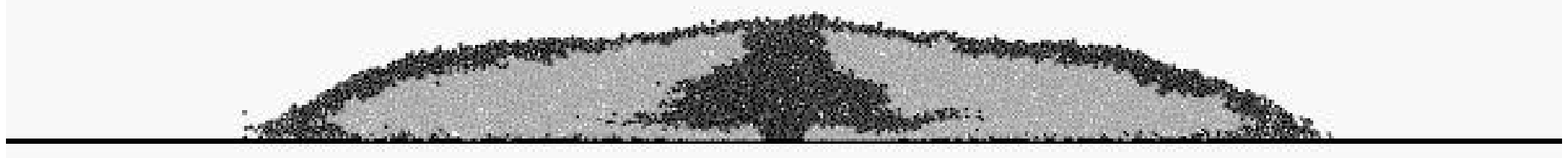}}
    \centerline{\includegraphics[width=0.99\linewidth]{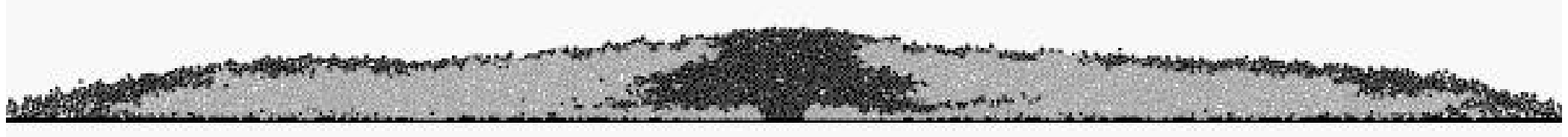}}
  \end{minipage}
\end{minipage}
\end{center}
\caption{Deformation of a collapsing column with $a = 9.1$, at $t/T_\infty=0$, $t/T_\infty=0.2$, $t/T_\infty=0.28$, $t/T_\infty=0.35$, $t/T_\infty=0.45$, and  $t/T_\infty=0.75$.}
%\caption{times 1, 15, 22, 28, 36, 50}
\label{Def}
\end{figure}

Basically, three successive stages can be identified in the history of a grain falling within the column. In a first step, its initial potential energy is converted into vertical motion, and if the initial height of the grain allows for it, it will be accelerated down to the bottom. There, in a second step, the grain will undergo collisions with the bottom plane or the surrounding grains, and its vertical motion will be converted into horizontal motion. In a third step, the grain eventually leaves the column base area and flow sideways. Of course, this process involves a collective dynamics of collisions and momentum lost and transfer, whose complexity makes any prediction of the trajectory of any grain dubious. For the same reason, a high initial potential energy is no individual guaranty, for a single grain, of a long sideways travel.\\
As an illustration of the complexity of the vertical to horizontal motion transfer, successive snapshots of the deformation of a collapsing column with $a\simeq 9.1$ are displayed in Fig~\ref{Def}. The grains initially situated at the margins of the column, in the central area, and at the top, are represented in black. In the course of time we observe that the grains traveling farther are not those initially at the top. On the contrary, grains which were at middle height or even lower are nearer to the spreading front area. This behaviour is also visible in the deformation of the inner black slice.\\

\begin{figure}
\centerline{\includegraphics[width=0.85\linewidth,angle = -90]{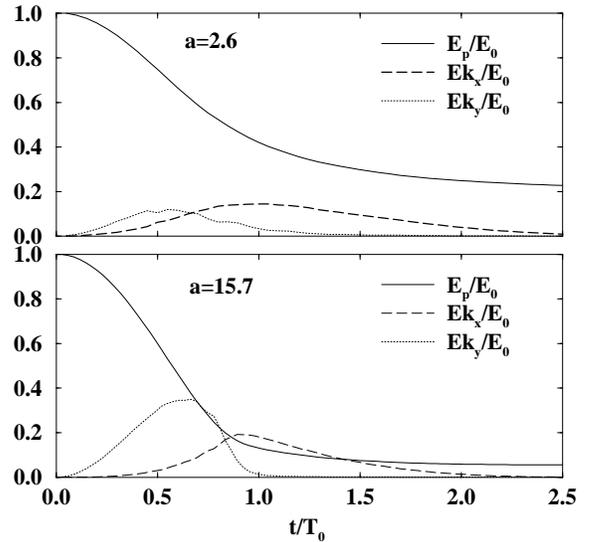}}
\caption{Time evolution of the potential energy $E_p$ (plain line), the vertical kinetic energy $E_{k_y}$ (dotted line) and the horizontal kinetic energy $E_{k_x}$ (dashed line) of the column for $a = 2.6$ (top) and $a = 15.7$ (bottom).}
\label{Eallt}
\end{figure}

The conversion of momentum from vertical to horizontal direction is likely to depend on the value of $a$. In Fig~\ref{Eallt}, we have represented for two collapsing columns with $a=2.6$ and $a=15.7$, the time evolution of the potential energy $E_p$, the vertical kinetic energy $E_{k_y}$ and the horizontal kinetic energy $E_{k_x}$ normalised by the initial potential energy $E_0$, with
\begin{eqnarray*}
E_p &=& \sum_{p=1}^{N_p} m_p g h_p,\\
E_{k_i} &=& \frac{1}{2} \sum_{p=1}^{N_p} m_p v_{p,i}^2,
\end{eqnarray*}
where $i=\{x,y\}$, $N_p$ is the total number of grains, $m_p$ their mass, $h_p$ their height and $\boldsymbol{v}_p$ their velocity.\\
From these graphs we first see that a higher proportion of the initial energy is dissipated by the column with $a = 15.7$, in agreement with the scaling of $\overline{H}_\infty/H_0$ in Figure~\ref{Hscale}, which is equivalent to the scaling of the final potential energy to the initial potential energy. Then, we observe that the conversion of energy from potential to kinetic in the vertical direction is much more efficient for $a = 15.7$. By contrast, a greater proportion of vertical kinetic energy is transferred into the horizontal direction for $a=2.6$. This suggests that the ability of the column to use its initial energy for spreading might depend on $a$.

\subsection{Conversion from Potential to Kinetic Energy}

For each value of $a$, we compute the kinetic energy of the system averaged over the total duration of the collapse $T_\infty$ in the vertical and horizontal direction:

\begin{equation}
\langle E_{k_i}\rangle = \frac{1}{T_\infty} \int_0^{T_\infty} \frac{1}{2}\sum_{p=1}{N_p}  m_p v_{p,i}^2 dt,\nonumber
\end{equation}
where $i=\{x,y\}$. To compare these values with predictions of the mean kinetic energy over the free fall time, and using the fact that $T_\infty \simeq 2.5 T_0$, with $T_0 = \sqrt(2H_0/g)$, the mean kinetic energies $\langle E_{k_x}\rangle$ and $\langle E_{k_x}\rangle$ are multiplied by a factor $2.25$ in the following discussion.

\begin{figure}
\begin{minipage}{0.98\linewidth}
\centerline{\includegraphics[angle=-90,width=0.9\linewidth]{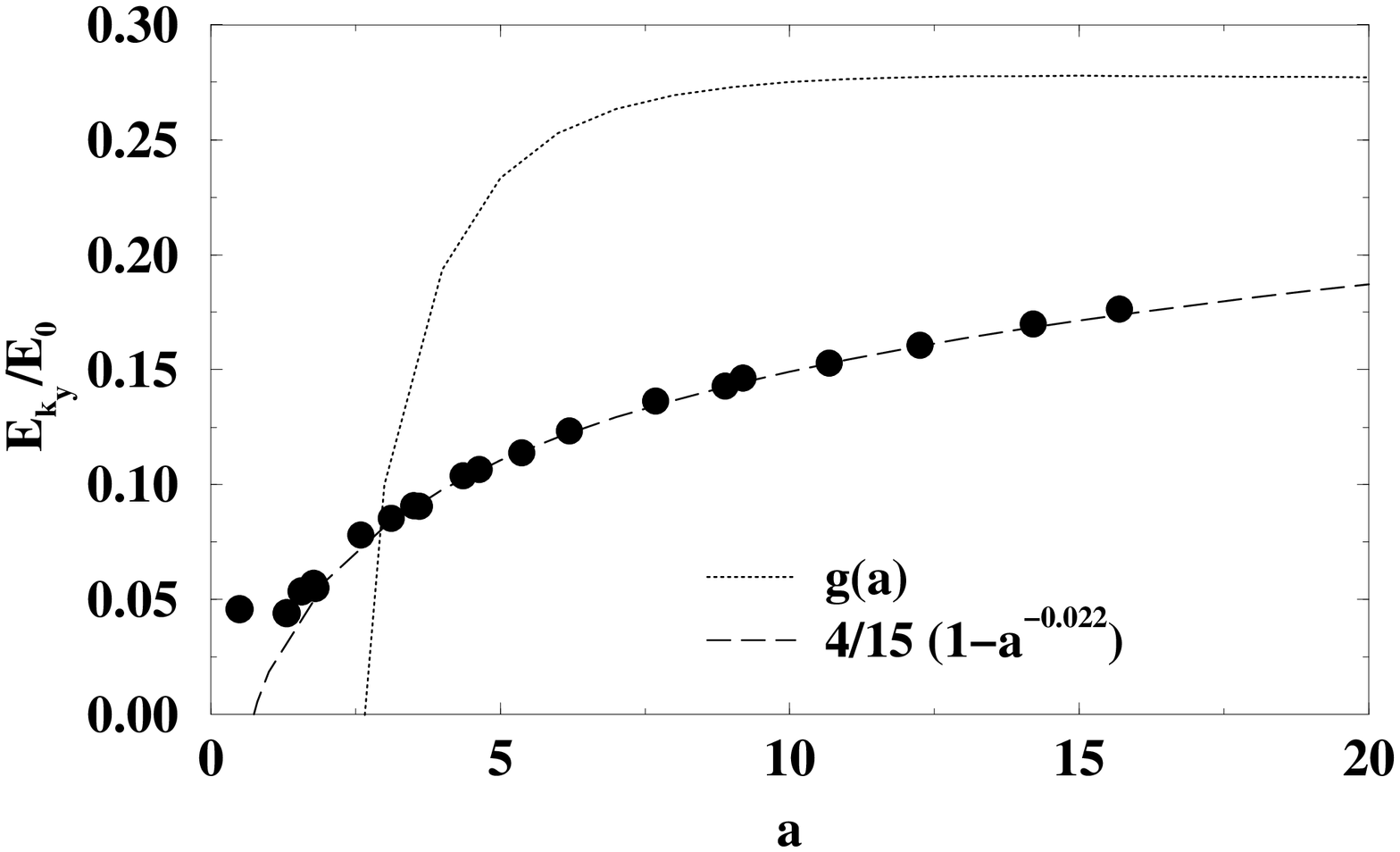}}
\caption{Vertical kinetic energy $\langle E_{k_y}\rangle$ averaged over the total duration of the collapse $T_\infty$ and normalised by $E_0$, as a function of the columns aspect ratio $a$.}
\label{Eky}
\end{minipage}
\vfill
\begin{minipage}{0.98\linewidth}
\centerline{\includegraphics[angle=-90,width=0.9\linewidth]{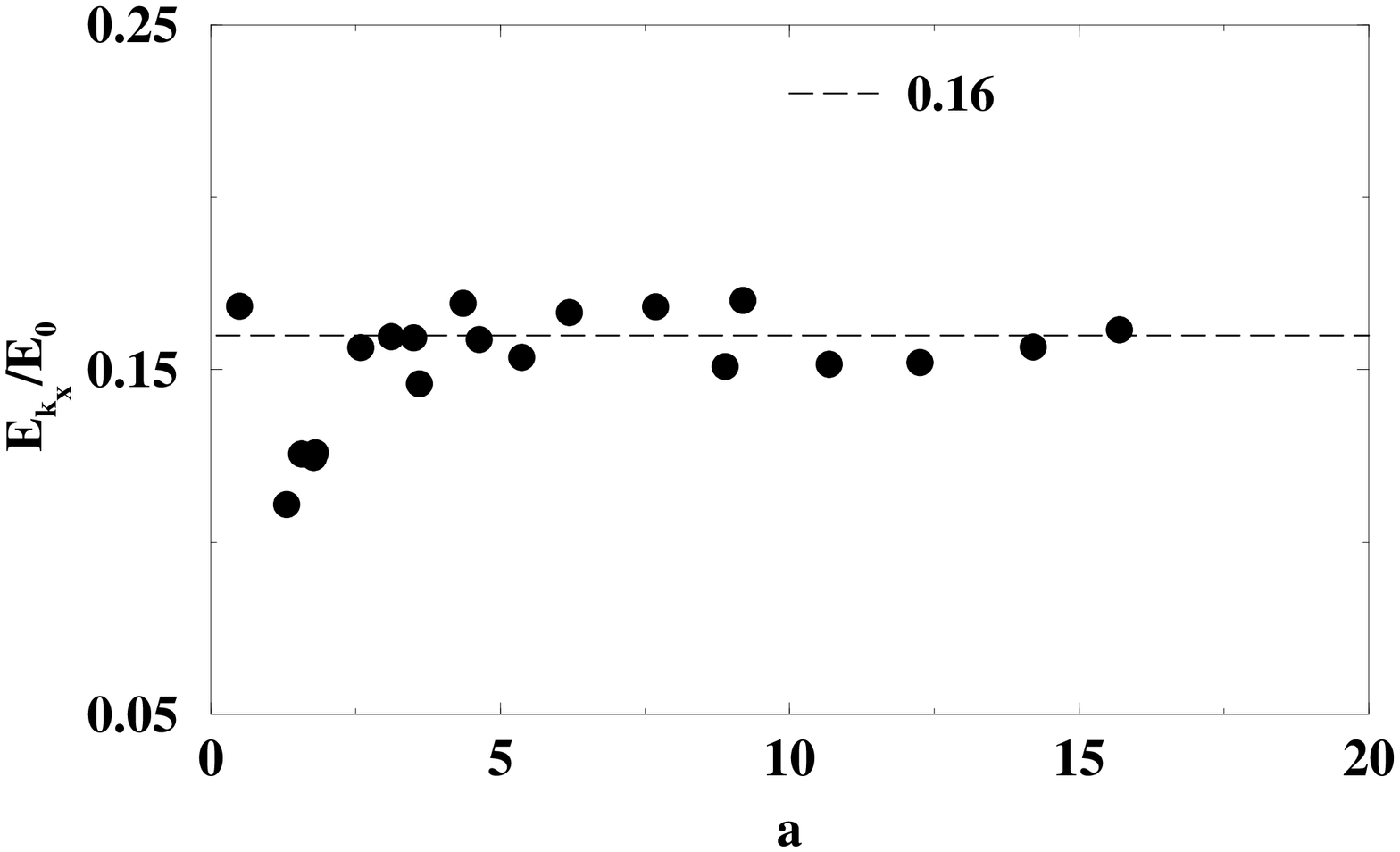}}
\caption{Horizontal kinetic energy $\langle E_{k_x}\rangle$ averaged over the total duration of the collapse $T_\infty$ and normalised by $aE_0$, as a function of the columns aspect ratio $a$.}
\label{Ekx}
\end{minipage}
\end{figure}

The mean kinetic energy of a column of height $H_0$ and radius $R_0$ submitted to free fall over a distance $H_0 -k R_0$, where $k$ is a constant, and averaged over the free fall time $T_f = \sqrt(2(H_0 -k R_0)/g)$ is given by:
$$\langle E_{k_y}\rangle = \frac{1}{T_f}\int_0^{T_f} \frac{1}{2} \rho_s R_0 (H_0 - \frac{1}{2}gt^2) g^2t^2 dt,$$
if we assume no influence of the grains which have already reached the bottom on the falling material.
This eventually leads to the relation
\begin{equation}
\frac{\langle E_{k_y}\rangle}{E_0} = (1-\frac{k}{a})(\frac{4}{15} + \frac{2}{5}\frac{k}{a}).
\label{Eqeky}
\end{equation}
For high values of the aspect ratio $a$ (or for $k=0$), the mean vertical kinetic energy is bounded by $4/15$. For intermediate values of $a$, $\langle E_{k_y}\rangle$ is expected to obey the relation~\ref{Eqeky} with $k=2.5$.
The mean vertical kinetic energy $\langle E_{k_y}\rangle$ renormalised by the initial energy $E_0$ is plotted in Fig~\ref{Eky} as a function of $a$. In the same graph is plotted the prediction $g(a) = (1-2.5/{a})({4}/{15} + {2}/{5}\times 2.5/{a})$ from the relation~\ref{Eqeky}. We observe that the simulations points and the prediction are very unlike, presumably because of the collective dynamics of the grains reaching the bottom of the column. However, following the prediction, the evolution of the mean vertical kinetic energy shows an increase (though much slower than would be expected) towards the limit value $4/15$, which appears in the following approximation of  $\langle E_{k_y}\rangle$:

\begin{equation}
\frac{\langle E_{k_y}\rangle}{E_0} \simeq \frac{4}{15}(1- a^{-0.022}).\nonumber
\end{equation}

We define the coefficient of restitution $\rho_y$ characterising the transfer of potential energy into vertical motion, so that 
$$\langle E_{k_y}\rangle = \rho_y E_0,$$ 
where $\rho_y = {4}/{15}(1- a^{-0.022})$ is an increasing function of $a$. For $a \rightarrow \infty$, $\rho_y \rightarrow 4/15$, namely the potential energy is converted into vertical motion following the ideal free fall case. For smaller values of $a$ however, there exists a transient regime during which the vertical kinetic energy increases with $a$. Similar transient behaviour in the dynamics of the sideways flow is discussed in next section.\\

The horizontal kinetic energy $\langle E_{k_x}\rangle$, normalised by $E_0$, is plotted in Fig~\ref{Ekx} as a function of $a$. We observe 
\begin{equation}
\frac{\langle E_{k_x}\rangle}{E_0} \simeq 0.16,\nonumber
\end{equation}
 for $a\gtrsim 2.5$. As preceedently we define the coefficient of restitution $\rho_x$ characterising the transfer of potential energy into horizontal motion:
$$\langle E_{k_x}\rangle = \rho_x E_0,$$ 
with $\rho_x = 0.16$. Surprisingly, in spite of the complexity of the process taking place at the bottom of the column, the energy available for the horizontal motion is simply proportional to the initial potential energy, and only implies the introduction of a constant effective coefficient of restitution. Hence we can no longer suspect the dissipation process at the bottom of the column to be at the origin of the divergence with the simple friction model described in equation~\ref{fric2}. 
We have seen in section~\ref{expe} that the scaling laws implied that the relation $ m_0gH_0 = \mu_e m_0gR_\infty$ is not satisfied. Since $\langle E_{k_x}\rangle = \rho_x E_0$, we cannot have $\langle E_{k_x}\rangle =  \mu_e m_0gR_\infty$ either, which would lead to $R_\infty = 0.16/\mu_e H_0$. We may thus suppose that the assumption of a basal friction controlling the energy dissipation in the sideways flow is wrong. This aspect is tackled in the next section.\\
Computing $E_{k_{y,max}}$ and $E_{k_{x,max}}$ the maximum kinetic energy in the vertical and horizontal direction during the collapse, the evolution of these two quantities as a function of $a$ is similar to the evolution of the corresponding mean quantities observed in Figure~\ref{Eky} and~\ref{Ekx}. The maximum values are about three times the mean values, as could be expected from the time evolution of $E_{k_y}$ and  $E_{k_x}$ shown in Figure~\ref{Eallt}.

Characterizing the ability of the columns to convert vertical kinetic energy into horizontal kinetic energy by the ratio $\rho_x/\rho_y$, we obtain a decreasing function of the aspect ratio tending towards $\rho_x$, suggesting again the existence of a transient regime in the dynamics of spreading.

\subsection{Basal Friction}

\begin{figure}
\centerline{\includegraphics[width=0.85\linewidth]{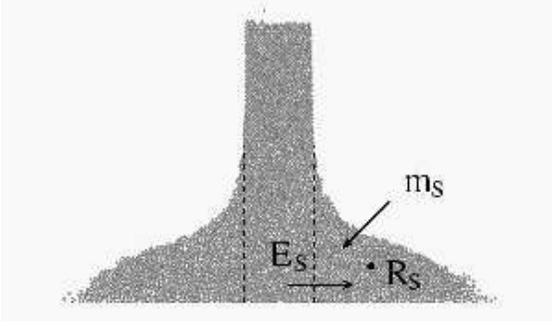}}
\caption{During the collapse, the exact amount of energy $E_S$ and the exact mass of grains $m_S$ crossing the section situated at $-R_0$ and $R_0$ are evaluated. The center of mass of the sideways flow is in $R_S$.}
\label{schemeA}
\end{figure}

The basal friction is the dissipation mechanism most often postulated for the dense flow of model granular media, and is indeed proven to be a relevant description~\cite[see][]{pouliquen99,pouliquen02}. It implies the definition of an effective coefficient of friction $\mu_e$, which is a mean phenomenological translation of the more complex dissipation process taking place at smaller scale through collisions and contact friction. \\
To check whether basal friction is a satisfactory description of the dissipation occurring within the sideways flow, we compare the exact amount of energy made available for the flow with the work done by the mass of grains over the distance they eventually run. Therefore, we define in the collapse two vertical sections $S$ situated at $-R_0$ and $R_0$, and delimiting the initial position of the mass of grains and the actual horizontal flow, as showned in Fig~\ref{schemeA}. Integrating over the total duration of the collapse $T_\infty$, we compute the exact amount of energy $E_S$ crossing the sections $S$, and thus taking part to the flow, and the exact mass of grains $m_S$ going through $S$. The energy $E_S$ is computed from both potential and kinetic energies of the grains. In the case of a simple friction process in the sideways flow, we should observe, independently of the dynamics occurring within the interval $[-R_0,R_0]$,
$$E_S = \mu_e m_S g (R_S-R_0),$$
where $R_S$ is the final distance run by the mass of grains $m_S$, and is taken  equal to the final position of the center of mass of $m_S$.\\

\begin{figure}
\centerline{\includegraphics[width=0.9\linewidth,angle = -90]{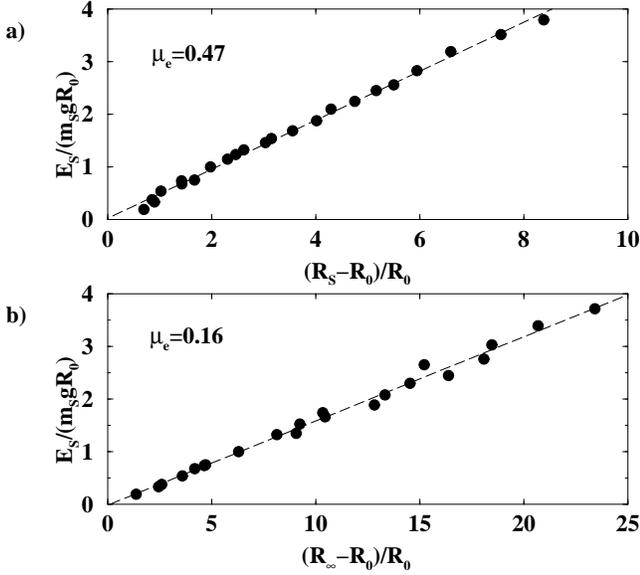}}
\caption{Energy available for the flow $E_S$ renormalised by the work of the flowing mass $m_S$ over $R_0$ as a function of the renormalised distance run by the center of mass of the flowing grains $(R_S-R_0)/R_0$ (graph a), and as a function of the renormalised runout distance of the flow $(R_\infty-R_0)/R_0$ (graph b)}
\label{FrotS}
\end{figure}

The energy $E_S$ normalised by the work of the mass $m_S$ over a distance $R_0$ is plotted as a function of the adimensional flowing distance $(R_S-R_0)/R_0$ in Figure~\ref{FrotS}a for all the collapse experiments. We obtain a neat linear dependance establishing that energy dissipation in the flow is very well approximated by basal friction. The value of the effective coefficient of friction $\mu_e$ is given by the slope of the linear relation and is found to be $\mu_e \simeq 0.47$.\\

If we assume that all the grains travel the final runout distance $R_\infty -R_0$, we again obtain a linear dependence, as seen in Figure~\ref{FrotS}b, but with a much smaller effective coefficient of friction $\mu_e \simeq 0.16$ expressing the maximum mobility of the flow. In any case, the following relation is verified:
\begin{equation}
E_S = \mu_e m_S g (R_\infty-R_0).
\label{frot1}
\end{equation}

\begin{figure}
\centerline{\includegraphics[width=0.9\linewidth,angle = -90]{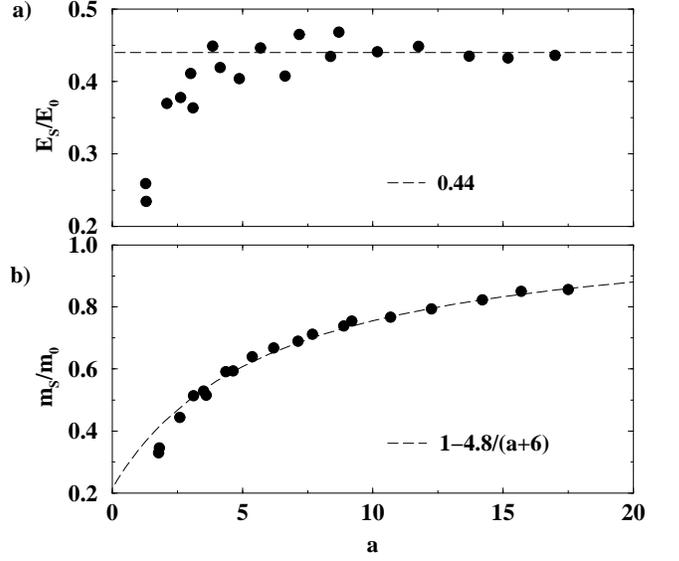}}
\caption{Energy available for the flow $E_S$ renormalised by the initial potential energy $E_0$ as a function of $a$ (graph a), and mass of grains $m_S$ actually taking part to the flow renormalised by the initial the total mass of grain $m_0$ as a function of $a$ (graph b).}
\label{EsMs}
\end{figure}

In Figure~\ref{EsMs}a, the plot of $E_S/E_0$ as a function of $a$ shows that as soon as $a \gtrsim 3$, $E_S/E_0 \simeq 0.44$. As could be inferred from the evolution of the mean horizontal energy (previous section), the energy available for the flow is simply proportional to the initial potential energy. The relation~\ref{frot1} can be rewritten:
\begin{equation}
0.44 E_0 \simeq \mu_e m_S g (R_\infty-R_0),
\label{frot2}
\end{equation}
or equivalently,
\begin{equation}
0.44 m_0 g H_0 \simeq \mu_e m_S g (R_\infty-R_0).
\label{frot3}
\end{equation}

Finally, this leads to the following expression for $a\gtrsim 3$:
$$ \frac{(R_\infty-R_0)}{R_0}\propto a \frac{m_S}{m_0}.$$
This relation suggests that the disagreement between the scalings observed experimentally and the simple friction model (see expression~\ref{fric2}) might rest in the definition of the mass of grains actually flowing.\\

The renormalised mass of grains actually crossing the section $S$ and taking part to the flow $m_S/m_0$ is plotted in Figure~\ref{EsMs}b as a function of $a$. There exist a function $f$ of the aspect ratio such as
$$\frac{m_S}{m_0} \simeq 1-f(a),$$
 with $f(a) \rightarrow 0$ when $a \rightarrow \infty$. No argument for the empirical fit $f(a) = 4.8 (6 + a)^{-1}$ is proposed; the main point of it is that it captures a major aspect of the flow phenomenology: the ejection of grains sideways. For small values of $a$, a small fraction of grains flows sideways, and most of the mass remains trapped in the interval $[-R_0,R_0]$, at the bottom of the initial column. The fraction of flowing grains increases with $a$. When $a$ becomes high, this fraction tends towards 1, as can be seen in Fig~\ref{Wave} where the mass flowing sideways in represented for $a = 70$. The increase of the proportion of mass actually taking part to the flow can be related to the increase of vertical kinetic energy relatively to the initial energy (see Fig~\ref{Eky}).\\
The increase of mass flowing sideways causes the friction dissipation process to be more efficient; while $a$ increases, the same fraction of initial potential energy $E_0$ is driving more mass against friction. This additional dissipation may well explain why an exponent lower than 1 appears in the scaling law. \\
This progressive increase of the flowing mass, bounded by the initial mass $m_0$, also suggests that the dynamics of the collapse, in the range of $a$ experimentally investigated, is transient. 

%---------------------------------------------------------
\section{New Scalings for a Transient Regime}
\label{new}

\begin{figure}
\begin{minipage}{0.99\linewidth}
\centerline{\includegraphics[width=0.9\linewidth,angle = -90]{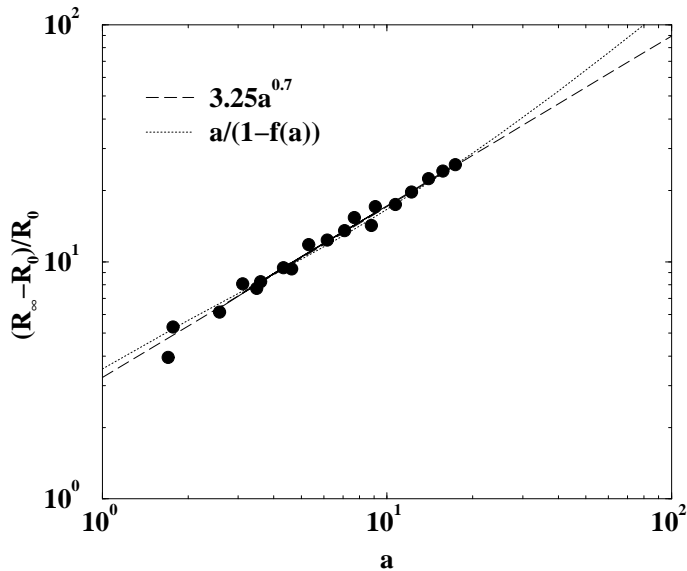}}
\caption{Final renormalised runout distance $(R_\infty-R_0)/R_0$ as a function of  $a$. Two different approximations are plotted: the power-law fit in dashed line, and the function $a/(1-f(a))$ in dotted line. }
\label{Rnew2}
\end{minipage}
\end{figure}

In the range of aspect ratios $a$ investigated in this present work, as well as in preceedent experimental work~\cite[see][]{lube04a,lajeunesse04,balmforth04}, the collapses of the columns mainly differ in the mass of grains ejected sideways. The higher the value of $a$, the greater the proportion of grains ejected. Because of this difference, the propagating flows do not involve the same proportion of the initial mass of grains. For increasing values of $a$, the increase of mass will cause the work done by the flow to be more efficient, affecting thus the runout distance.\\
However, the mass of grains flowing tends towards the initial mass $m_0$ when $a$ increases, so we anticipate the dependance of the dissipation process on $a$ to vanish. In this limit, the amount of energy dissipated by the flow would only be dependent on the runout distance.\\ 
The dependance of the proportion of mass actually flowing $m_S/m_0$ on $a$ shown in Figure~\ref{EsMs}b reflects this behaviour and the asymptotic evolution towards a new regime. However the scaling laws displayed in Figure~\ref{Rscale} and Figure~\ref{Hscale} do not at all express the existence of this transition.\\
On the basis of our observations of the phenomenology of the sideways flow, we can thus suggest that there exists a function $f(a)$, verifying $f(a) \rightarrow 0$ when $a\rightarrow \infty$, such as:
\begin{equation}
\frac{R_\infty -R_0}{R_0} \propto \frac{a}{(1-f(a))}.
\end{equation}
We have represented in Figure~\ref{Rnew2} the evolution of $(R_\infty -R_0)/{R_0}$ with $a$, the power law approximation $3.25 a^{0.7}$, and the empirical fit ${a}/(1-f(a))$, where $f(a) = 4.8 (6 + a)^{-1}$ describes the grains ejection process (see section above). We observe that in the range of aspect ratios investigated, the last choice is as acceptable as a power-law dependence.\\

No argument on the form of the function $f(a)$ will be discussed at this stage.  Basically, it represents the additional dissipation entailed by the increase of the proportion of mass flowing, while the proportion of energy available for the flow is constant (at least for $a\gtrsim 3$ in our simulations). The choice of an approximation of the form ${a}/(1-f(a))$ for the runout distance has two implications:
\begin{enumerate}
\item For large $a$, the runout distance will eventually increase like the height of the column, as expected in a simple friction dynamics.  
\item It suggests that the power-law dependance might be fortuitous.
\end{enumerate}

In the absence of a comprehensive model explaining either one or the other approximation, the proposition discussed here remains purely speculative. However, it matches nicely the numerical results, and provides a qualitative explanation for the non linear behaviour of the runout distance with $a$. Finally, it suggests that the key point of the collapse problem lies in the dynamics of ejection of the mass from the initial column itself, rather than in the characteristics of the sideways flow.

% coefficient of friction independant of the mass flowing

\section{Summary and Conclusion}
\label{conclu}

We have numerically investigated the collapse and the spreading of two-dimensional columns of grains onto a vertical plane using the Contact Dynamics method. This approach allows for a detailed analysis of the dynamics of the collapse taking into account the individual grains energy and trajectory. Our results are generally in good agreement with previous experimental work carried out in quasi-2D configurations~\cite[see][]{lube04b,balmforth04}.\\
The collapse is first described in terms of the shape of the final deposit, and more specifically in terms of runout distance. A power law dependence of the renormalized runout distance with the initial aspect ratio of the columns is found for high aspect ratios, and a linear dependence is found for low aspect ratios. These scalings, experimentally observed by previous authors, are incompatible with a simple friction model of the collapse dynamics.\\
We show that the collapse is driven by the free fall of the column for high enough aspect ratios. The existence, or absence, of free fall dynamics, can explain the existence of two different scaling laws for the runout distance depending on the aspect ratio. The propagation of the front involves a constant velocity phase, followed by a deceleration phase of significant contribution to the runout distance. The analysis of the mean kinetic energy of the grains shows that the dissipation process occurring at the bottom of the column can be simply described by a constant coefficient of restitution. In particular, the energy avalaible for the sideways flow is simply proportional to the initial potential energy. The detailed analysis of the energy dissipated in the sideways flow and the work of the flowing mass clearly establishes that friction is a good approximation of the dissipation process. Finally, we point at the dynamics of mass ejection sideways during the column collapse as playing a predominant role in the spreading dynamics, and as being responsible for the non-linear behaviour of the renormalised runout distance. This allows us to suggest that the scaling laws preceedently discussed for the runout distance are fortuitous, and should no longer apply when the aspect ratio increases. A new empirical fit is proposed, which is compatible with a friction model. \\

This conclusion has different implications:
\begin{enumerate}
\item Mass ejection sideways is a mechanism strongly dependent on the geometry of the collapse. In particular, we expect its effects to be more important in a 2D configuration, than in the case of an axisymmetric collapse. This might be at the origin of the difference is the scaling laws for the runout distance observed in 2D (or quasi-2D) and axisymmetric configurations. In the limit of high aspect ratios however, the mass ejection tends towards the ejection of the totality of the initial mass. In that limit, differences should no longer be observed between the 2D and the axisymmetric configuration. In any case, our results suggest that an analytical expression of the runout distance should account for the process of the ejection of grains at the bottom of the collapsing column. 
\item  The runout distance appears to be strongly dependent on the fall dynamics and not only on the effective flow properties, namely effective basal friction in our case.  Although high aspect ratios are difficult to find in natural context, many rock falls or slope destabilisations involve a strong acceleration (and possibly free fall), which is a key aspect of the material ejection. In a geophysical perspective, this suggests that the mobility of a natural flow, usually defined as the ratio of the runout distance to the initial height of the material, is related to the early dynamics of the mass release as well as to the flowing properties of the material.
\item Since the sideways flow undergoes a simple basal friction dissipation process, its modelling using shallow water approaches is possible, but not straightforward. Indeed, a difficulty lies in the description of the initial conditions represented by the vertical column collapse, which escapes, intrinsically, the shallow water assumptions. The issue is to achieve a correct description of the mass flux with a limitation of the energy induced by the column fall. The column collapse dynamics, until now correctly reproduced only for low aspect ratios~\cite[see][]{mangeney04,kerswell04}, could also be recovered for high aspect ratios~\cite[see][]{larrieu04}.      
\end{enumerate}
The influence of the material properties (inter-grains friction $\mu$ and restitution at collision $\rho$) on the overall dynamics of the collapse and the spreading will be the subject of further works.\\

This work was supported by the Marie Curie European Fellowship FP6 program.

%\bibliographystyle{jfm}

%\bibliography{Jet2jfm}

\end{document}